\documentclass[aps,10pt,prd,onecolumn,groupedaddress,floatfix,longbibliography,superscriptaddress,notitlepage,tikz]{revtex4-2}

\usepackage{amsmath}
\usepackage{amssymb}
\usepackage{amsfonts}
\usepackage{bm}
\usepackage{times,float}
\usepackage{graphicx}
\usepackage[dvipsnames,svgnames]{xcolor}
\usepackage{hyperref}
\hypersetup{colorlinks=true, linkcolor=NavyBlue, citecolor=PineGreen,urlcolor=cyan}
\usepackage{physics}
\graphicspath{{FIGURES/}}
\usepackage{mathrsfs}
\usepackage{subcaption}
\usepackage{tikz}
\usetikzlibrary{arrows.meta}

\newcommand{\partd}[3][1]{%
  \ifnum#1=1
    \frac{\partial #2}{\partial #3}%
  \else
    \frac{\partial^{#1}#2}{\partial #3^{#1}}%
  \fi
}

\begin{document}

\title{Electromagnetic response of two interacting topological insulator spheres in external fields}

\author{J. Cornejo Gómez}
\affiliation{Instituto de Ciencias Nucleares, Universidad Nacional Aut\'{o}noma de M\'{e}xico, 04510 Ciudad de M\'{e}xico, M\'{e}xico}

\author{M. Ibarra-Meneses}
\affiliation{Instituto de Ciencias Nucleares, Universidad Nacional Aut\'{o}noma de M\'{e}xico, 04510 Ciudad de M\'{e}xico, M\'{e}xico}

\author{L. Medel Onofre}
\address{Instituto de Ciencias Nucleares, Universidad Nacional Aut\'{o}noma de M\'{e}xico, 04510 Ciudad de M\'{e}xico, M\'{e}xico}

\author{A. Mart\'{i}n-Ruiz}
\email{alberto.martin@nucleares.unam.mx}
\affiliation{Instituto de Ciencias Nucleares, Universidad Nacional Aut\'{o}noma de M\'{e}xico, 04510 Ciudad de M\'{e}xico, M\'{e}xico}

\begin{abstract}
We study the static electromagnetic response of two spherical topological insulators embedded in a dielectric medium and subjected to a uniform external electric field. The gapped surface states are described by a piecewise constant axion field, which induces a topological magnetoelectric coupling localized at the spherical interfaces. {More generally, the same formalism applies to isotropic magnetoelectric media characterized by an effective scalar magnetoelectric response.} The electrostatic problem is solved at zeroth order using bispherical coordinates, allowing for an exact treatment of both parallel and perpendicular orientations of the external field relative to the center-to-center axis. The resulting mode expansions are determined by three-term recurrence relations, which are solved perturbatively for nonoverlapping spheres. The { magnetoelectric}-induced response is then computed to leading order in the fine-structure constant {(or, more generally, in the effective coupling strength)}. The induced sources are purely interfacial and generate distinct magnetostatic field configurations in the parallel and perpendicular geometries. Closed-form series representations for the induced vector potential and magnetic field are obtained in terms of the zeroth-order electrostatic coefficients. These results provide an analytically controlled description of {interaction-induced magnetostatics in coupled spherical magnetoelectric systems}.
\end{abstract}

\maketitle


\section{Introduction}

Topological phases of matter have reshaped our understanding of electronic systems by revealing that global, topological properties of Bloch wave functions can determine robust and quantized physical responses \cite{thouless_1982, hasan_kane_2010, qi_zhang_2011}. In three dimensions, topological insulators represent a paradigmatic example: they exhibit an insulating bulk gap coexisting with metallic surface states protected by time-reversal symmetry and characterized by an odd number of Dirac cones \cite{fu_kane_mele_2007, moore_balents_2007, roy_2009}. Beyond bulk crystals, increasing attention has been devoted to finite and mesoscopic realizations of topological insulators, including thin films \cite{okamoto_imura_2014}, nanowires \cite{governale_bhandari_2020}, and nanoparticles, where geometry, confinement, and boundary conditions qualitatively modify the electronic and electromagnetic response \cite{cook_nielsen_2023}.

A defining macroscopic consequence of the nontrivial topology of three-dimensional topological insulators is the topological magnetoelectric effect. At low energies, this response is described by axion electrodynamics, in which Maxwell's theory is supplemented by a term proportional to $\theta \, \mathbf{E} \cdot \mathbf{B}$ \cite{qi_hughes_zhang_2008, essin_magnetoelectric_2009}. For strong topological insulators, the axion angle is quantized as $\theta = \pm \pi$, while trivial insulators correspond to $\theta = 0$. When time-reversal symmetry is broken at the surface, for instance by magnetic coatings or exchange coupling, this term leads to quantized surface Hall conductivities and unconventional electromagnetic phenomena such as image magnetic monopoles \cite{qi_monopole_2009,PhysRevLett.103.171601,PhysRevD.92.125015} and repulsive Casimir effect \cite{PhysRevLett.106.020403, PhysRevB.84.045119, Martín-Ruiz_2016}. Importantly, when $\theta$ is piecewise constant, the axion contribution vanishes in the bulk and manifests itself exclusively through interfacial sources, making finite geometries essential for observable effects.  {While our discussion is framed in the language of topological insulators and axion electrodynamics, the mathematical structure developed here applies more generally to isotropic magnetoelectric media described by a scalar magnetoelectric coupling. In this broader context, the coupling need not be quantized nor of topological origin. Materials with sufficiently high crystallographic symmetry, including certain non-collinear antiferromagnets exhibiting approximately isotropic magnetoelectric response \cite{Spaldin_2008, Malashevich_2010, PhysRevLett.109.197203}, may therefore provide a more experimentally realistic platform for observing related interaction-induced electromagnetic effects.}

From the standpoint of classical electrodynamics, the response of dielectric bodies to external electromagnetic fields constitutes a paradigmatic boundary-value problem with a long and well-established history \cite{jackson_classical,morse_feshbach}.  Among the simplest nontrivial geometries, the system of two interacting spheres occupies a central role in electrostatics \cite{PhysRevB.13.4320,10.1063/1.343737,10.1098/rspa.2012.0133} and electromagnetic scattering theory \cite{leichner_two_spheres,batool_two_spheres_2022}, as it captures multiple-scattering processes, near-field coupling, and collective polarization effects that are absent in isolated-particle descriptions. Such two-body configurations arise naturally in a wide variety of mesoscopic and composite systems.  In nanoparticle assemblies and colloidal suspensions, electromagnetic interactions between nearby inclusions lead to collective optical and electrostatic responses that cannot be accounted for within single-particle approximations \cite{prodan_plasmonics,markel_dimer_review,brongersma_plasmon_review,10.1063/1.4961091}.  At the minimal level, plasmonic and dielectric dimers provide the simplest setting in which near-field coupling, mode hybridization, and symmetry-induced selection rules can be analyzed in a controlled and analytically transparent manner \cite{prodan_plasmonics,nordlander_hybridization,leichner_two_spheres}. More generally, ordered or disordered arrays of closely spaced particles form the building blocks of artificial metamaterials, whose effective electromagnetic properties are governed primarily by multiple-scattering effects and inter-particle interactions rather than by the response of individual constituents \cite{pendry_metamaterials,shalaev_metamaterials_review,markel_effective_media}. For these reasons, the two-sphere problem provides a paradigmatic and analytically tractable framework for exploring interaction-induced modifications of electromagnetic response in structured and topologically nontrivial media. Closely related two-body geometries have also been extensively investigated. In particular, the sphere-plane configuration has attracted sustained attention in electrostatics  since  it provides a realistic model for particle-surface interactions and probe-sample  coupling. From a geometric standpoint, this setup can be understood as a limiting  case of the two-sphere system in which the radius of one sphere tends to infinity, so that many conceptual and methodological aspects are shared between both problems \cite{hudlet_epjb_1998, 10.1063/1.4862897,7862209}. When formulated in bispherical coordinates, the two-sphere problem in the presence of a field admits an exact separation of variables and allows for analytically controlled solutions \cite{morse_feshbach}.

Motivated by these developments, in this work we study the static electromagnetic response of two interacting topological insulator spheres embedded in a dielectric medium and subjected to a uniform external electric field. We first solve the classical electrostatic problem exactly at zeroth order using bispherical coordinates, obtaining mode expansions governed by three-term recurrence relations. These relations are solved perturbatively for nonoverlapping spheres, providing an analytically transparent description of the interaction-induced corrections. Related two-body configurations involving topological media (such as a metallic sphere near a topological insulator surface) have been previously analyzed in the context of axion electrodynamics and induced magnetoelectric effects \cite{PhysRevA.100.042124}, highlighting the measurable consequences of interfacial topological couplings. Building on the exact two-sphere electrostatic solution, we then compute the leading axion-induced electromagnetic response to first order in the fine-structure constant. Because the axion field is piecewise constant, the induced sources are purely interfacial and generate distinct magnetostatic field configurations in the parallel and perpendicular geometries. Our results provide an analytically controlled framework for axion-induced magnetostatics in interacting finite systems and suggest experimentally accessible signatures of topological nontriviality in multi-sphere geometries.

The remainder of this article is organized as follows. In Sec.~\ref{Axion_ED_TIs} we briefly review the axion electrodynamics framework appropriate for three-dimensional topological insulators, emphasizing the form of the modified Maxwell equations and the role of piecewise constant axion fields. In Sec.~\ref{problem_section} we introduce the general formulation of the problem, including the geometrical setup, the use of bispherical coordinates, and the perturbative expansion scheme employed to solve the axion-modified field equations. Section~\ref{zeroth_order_solution} is devoted to the electrostatic problem of two dielectric spheres subjected to a uniform external electric field, which serves as the seed solution for the subsequent magnetoelectric analysis. In this section, both parallel and perpendicular orientations of the applied field are treated, and the resulting mode expansions are determined through three-term recurrence relations that are solved perturbatively for nonoverlapping spheres. In Sec.~\ref{induced_magnetic_field_section} we compute the leading axion-induced magnetic response generated by the topological magnetoelectric effect, again for both parallel and perpendicular configurations, and obtain explicit series representations for the induced vector potential and magnetic field. Finally, in Sec.~\ref{results_discussion_section} we summarize our main results and discuss possible extensions and implications of our work.

\section{Axion electrodynamics of topological insulators} \label{Axion_ED_TIs}

The macroscopic electromagnetic response of 3D TIs, independently of microscopic details, is described by an effective field theory that extends Maxwell electrodynamics by a topological axion term. In SI units, the corresponding action reads \cite{qi_hughes_zhang_2008, essin_magnetoelectric_2009}
\begin{align}
S = \int d ^{4} x \, \left[ \frac{1}{2} \left( \epsilon \mathbf{E} ^{2} - \frac{1}{\mu} \mathbf{B} ^{2} \right) + \frac{\alpha}{\pi} \sqrt{\frac{\epsilon _{0}}{\mu _{0}}} \, \theta \, \mathbf{E} \cdot \mathbf{B} \right] , \label{Action}
\end{align}
where $\alpha = e ^{2}/( 4 \pi \epsilon _{0} \hbar c )$ is the fine-structure constant, $\epsilon$ and $\mu$ denote the permittivity and permeability of the material, respectively, and $\theta$ is the topological magnetoelectric polarizability (axion field).

Time-reversal symmetry constrains $\theta$ to the values $\theta = 0$ or $\pi$ (mod $2\pi$). As a result, the axion term does not modify Maxwell's equations in the bulk of a homogeneous TI. Its physical consequences arise only at interfaces where $\theta$ changes discontinuously, such as between a TI and a conventional insulator. When the surface states are gapped by a TR-breaking perturbation (induced, for instance, by magnetic doping or by the application of an external static magnetic field) the system becomes a full insulator characterized by a quantized value $\theta = \pm \pi$.

Varying the action~\eqref{Action} yields Maxwell's equations in matter supplemented by modified constitutive relations,
\begin{align}
\mathbf{D} &= \epsilon \mathbf{E}
+ \alpha \frac{\theta}{\pi} \sqrt{\frac{\epsilon_{0}}{\mu_{0}}} \, \mathbf{B},
\qquad
\mathbf{H} = \frac{1}{\mu} \mathbf{B}
- \alpha \frac{\theta}{\pi} \sqrt{\frac{\epsilon_{0}}{\mu_{0}}} \, \mathbf{E},
\label{ConstEquations}
\end{align}
which encode the topological magnetoelectric effect: an electric field induces a magnetic polarization, while a magnetic field induces an electric polarization \cite{qi_hughes_zhang_2008, essin_magnetoelectric_2009}.

In the static regime relevant for this work, Maxwell's equations reduce to
\begin{align}
\nabla \cdot \left[ \epsilon(\mathbf{r}) \mathbf{E}(\mathbf{r}) \right]
&= \rho(\mathbf{r})
- \frac{\alpha}{\pi} \sqrt{\frac{\epsilon_{0}}{\mu_{0}}}
\, \nabla\theta(\mathbf{r}) \cdot \mathbf{B}(\mathbf{r}),
\label{GaussE_static} \\
\nabla \times \left[ \mu^{-1}(\mathbf{r}) \mathbf{B}(\mathbf{r}) \right]
&= \frac{\alpha}{\pi} \sqrt{\frac{\epsilon_{0}}{\mu_{0}}}
\, \nabla\theta(\mathbf{r}) \times \mathbf{E}(\mathbf{r}),
\label{Ampere_static}
\end{align}
supplemented by the homogeneous equations
\begin{align}
\nabla \cdot \mathbf{B} = 0, \qquad \nabla \times \mathbf{E} = \mathbf{0}.
\end{align}
Since $\theta$ is piecewise constant, its gradient vanishes everywhere except at material interfaces. For a TI region in contact with a conventional insulator across an interface $\Sigma$, one has
\begin{align}
\nabla\theta(\mathbf{r}) = (\theta_{2}-\theta_{1}) \, \delta(\Sigma)\,\hat{\mathbf{n}},
\end{align}
where $\hat{\mathbf{n}}$ is the outward unit normal to $\Sigma$. As a result, the axion term affects the electromagnetic response only through modified interfacial conditions, while the bulk fields obey the conventional electrostatic and magnetostatic equations. Static problems involving topological insulators can therefore be formulated as boundary-value problems closely analogous to classical dielectric systems, but augmented by topological magnetoelectric couplings localized at the material interfaces \cite{PhysRevD.92.125015,PhysRevD.94.085019}.

{

\section{Exact solution for an isolated topological-insulator sphere}

Before addressing the interacting two-sphere geometry, it is instructive to consider the simpler case of a single isolated topological-insulator sphere in a uniform external electric field. Besides its exact solvability, this configuration provides a natural benchmark for the more general interacting problem studied below, allowing us to identify the basic magnetoelectric response induced by the axion coupling in the absence of inter-sphere interactions.

We consider a spherical topological insulator of radius $R$, characterized by constitutive parameters $(\epsilon_1,\mu_1,\theta)$, embedded in a topologically trivial dielectric medium $(\epsilon_2,\mu_2)$ with $\theta=0$, and subjected to a static uniform electric field $\mathbf E_0 = E_0 \hat{\mathbf z}$. Inside the sphere, the constitutive relations take the axion-electrodynamics form
\begin{align}
\mathbf D_1=\epsilon_1 \mathbf E_1+\gamma \mathbf B_1,
\qquad
\mathbf H_1=\frac{1}{\mu_1}\mathbf B_1-\gamma \mathbf E_1,
\end{align}
where $\gamma \equiv \alpha  (\theta/\pi )\sqrt{\frac{\epsilon_0}{\mu_0}}$, while outside the sphere one has the conventional relations $\mathbf D_2=\epsilon_2 \mathbf E_2$ and $\mathbf H_2= \mathbf B_2 / \mu_2$.
In the static regime, both the electric and magnetic fields may be expressed in terms of scalar potentials satisfying Laplace’s equation in each region. Owing to axial symmetry, only the dipolar ($l=1$) sector contributes, so that the potentials take the form
\begin{align}
    \Phi_1 = -A r \cos\vartheta, \qquad \Phi_2 = -E_0 r\cos\vartheta + \frac{P\cos\vartheta}{r^2} , \qquad \Psi_1 = -C r \cos\vartheta, , \qquad \Psi_2 = \frac{M\cos\vartheta}{r^2} . 
\end{align}
Imposing the standard Maxwell boundary conditions at the spherical interface, including the axion-modified constitutive response inside the sphere, the coefficients can be determined exactly. The resulting interior fields are found to be uniform,
\begin{align}
\mathbf E_1=
\frac{3\epsilon_2}
{
\epsilon_1+2\epsilon_2
+
\tilde{\mu} \gamma^2 
} \, 
\mathbf E_0, \qquad \qquad
\mathbf H_1=
- \frac{\gamma \tilde{\mu}}{2 \mu_2}
\frac{3  \epsilon_2  }
{
\epsilon_1+2\epsilon_2
+
\tilde{\mu} \gamma^2 
} \, 
\mathbf E_0 ,
\end{align}
where $\tilde{\mu} = 2 \mu _{1} \mu _{2} / (\mu _{1} + 2 \mu _{2} ) $. Thus, the axion coupling induces a uniform magnetic response inside the topological sphere even in the absence of externally applied magnetic fields.

Outside the sphere, the electromagnetic response acquires the standard dipolar form. The electric field $\mathbf E_2$ consists of the superposition of the externally applied uniform field $\mathbf E_0$ and the field generated by an induced electric dipole moment $\mathbf p$. Similarly, the axion-induced magnetic response outside the sphere $\mathbf H_2$ corresponds to the field of an induced magnetic dipole $\mathbf m$. These dipolar moments are found to be
\begin{align}
\mathbf P = 4\pi \epsilon_2 R^3 \,  \frac{
\epsilon_1 - \epsilon_2
+\tilde{\mu} \gamma^2}
{
\epsilon_1+2\epsilon_2
+
\tilde{\mu} \gamma^2
}  \, \mathbf E_0 , \qquad \qquad \mathbf M =     \frac{\gamma \tilde{\mu}}{2 \mu_2}
\frac{3  \epsilon_2  }
{
\epsilon_1+2\epsilon_2
+
\tilde{\mu} \gamma^2 
}  4\pi   R^3\, \mathbf E_0 . 
\end{align}
Therefore, the exact solution exhibits the characteristic topological magnetoelectric effect in its simplest form: an externally applied electric field generates an induced magnetic dipolar response mediated entirely by the axion coupling at the spherical interface. This isolated-sphere solution provides the natural reference configuration against which the genuinely interaction-induced modifications of the two-sphere geometry can be assessed.

}

\section{Problem statement} \label{problem_section}

\begin{figure}
\includegraphics[scale=1]{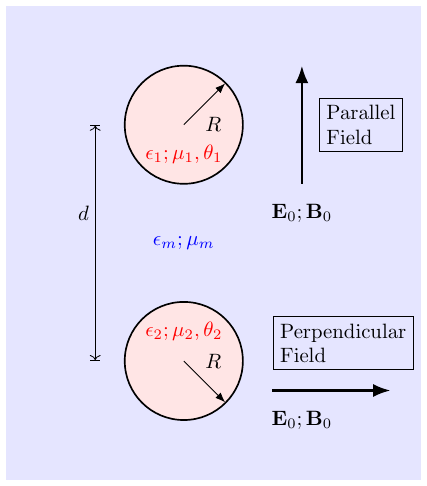}
\caption{
Schematic configuration of the system. Two spherical TIs of radius $R$, characterized by constitutive parameters $(\epsilon_1,\mu_1,\theta _1)$ and $(\epsilon_2,\mu_2 , \theta _2)$, are embedded in a homogeneous external medium with parameters $(\epsilon_m,\mu_m)$ and separated by a center-to-center distance $d$. The figure illustrates the two relevant orientations of the externally applied uniform fields, either electric ($\mathbf E_0$) or magnetic ($\mathbf B_0$): parallel to the inter-sphere axis (upper panel) and perpendicular to it (lower panel). These two geometries define the distinct configurations analyzed throughout this work.
} \label{configuration}
\end{figure}

We consider the static electromagnetic response of two spherical topological insulators immersed in a homogeneous dielectric medium and subjected to an externally applied uniform electromagnetic field, as illustrated in {Fig.~\ref{configuration}}. Each sphere has radius $R _{i}$ $(i=1,2)$ and is characterized by a dielectric constant $\epsilon _{i}$, magnetic permeability $\mu _{i}$, and a topological magnetoelectric polarizability $\theta _{i}$. The surrounding medium is a conventional insulator with dielectric constant $\epsilon _{m}$ and permeability $\mu _{m}$.

Since trivial insulators and known 3D TIs are typically nonmagnetic, we assume $\mu_{1}=\mu_{2}=\mu_{m}=\mu_{0}$. 
The spheres are identical and have the same radius $R$, and are separated by a center-to-center distance $d$. 
The system contains no free charges or currents, so that the electromagnetic response is entirely induced by the external fields and the material interfaces.

In the specific case of three-dimensional topological insulators, a nontrivial magnetoelectric response arises when the surface states of each sphere are gapped by a time-reversal-symmetry-breaking perturbation, leading to quantized values $\theta_i=\pm\pi$ inside the topological regions and $\theta=0$ in the surrounding dielectric. As a result, the axion contribution is confined to the spherical interfaces, while the electromagnetic fields in the bulk obey the conventional electrostatic and magnetostatic equations. {More generally, however, the present formalism is not restricted to quantized topological insulators, but applies to any isotropic magnetoelectric medium whose constitutive response can be described by an effective scalar magnetoelectric coupling.} We focus on the static response of the system to an externally applied uniform electric field, which can therefore be formulated as a boundary-value problem for Maxwell’s equations with magnetoelectric interfacial couplings.

\subsection{Geometry of the problem}

The presence of two interacting spherical bodies naturally suggests the use of a coordinate system adapted to spherical boundaries. We therefore employ the bispherical coordinate system, which is particularly well suited for boundary-value problems involving two spheres and has been extensively used in classical electrostatics and magnetostatics \cite{MoonSpencer1961,MorseFeshbach1953}. In this representation, the surfaces of both spheres are described by constant coordinate values, which simplifies the formulation of the problem.

The relation between the Cartesian $(x,y,z)$ and the bispherical $(\eta,\xi,\phi)$ coordinate systems is given by
\begin{align}
x = \frac{a\sin{\xi}\cos{\phi}}{\cosh{\eta}-\cos{\xi}}, \qquad
y = \frac{a\sin{\xi}\sin{\phi}}{\cosh{\eta}-\cos{\xi}}, \qquad
z = \frac{a\sinh{\eta}}{\cosh{\eta}-\cos{\xi}},
\label{bispherical-coordinates}
\end{align}
where $\phi$ is the azimuthal angle around the $z$ axis and $a>0$ sets the length scale of the coordinate system. The coordinate ranges are
\begin{align}
- \infty \leq \eta \leq + \infty, \qquad
0 \leq \xi \leq \pi, \qquad
0 \leq \phi \leq 2 \pi.
\end{align}
In this coordinate system, surfaces of constant $\eta$ represent nonintersecting spheres with centers on the $z$ axis. A surface $\eta = \eta _{0}$ corresponds to a sphere of radius
\begin{align}
R = \frac{a}{|\sinh{\eta_{0}}|},
\end{align}
whose center is located at
\begin{align}
z_{c} = a\,\coth{\eta_{0}}.
\end{align}
Positive (negative) values of $\eta _{0}$ correspond to spheres centered on the positive (negative) $z$ axis. The limiting surface $\eta = 0$ represents a sphere of infinite radius and therefore corresponds to the plane $z = 0$. Surfaces of constant $\xi$ form tori intersecting the $z$ axis, while surfaces of constant $\phi$ are half-planes bounded by that axis.

We consider two physical spherical topological insulators whose surfaces are described by $\eta=\eta_{1}>0$ and $\eta=\eta_{2}<0$, respectively, as shown in {Fig.~1}. Their radii are
\begin{align}
R_{1} = \frac{a}{\sinh{\eta_{1}}}, \qquad
R_{2} = \frac{a}{|\sinh{\eta_{2}}|},
\end{align}
and the distance between their centers is
\begin{align}
d = a\left(\coth{\eta_{1}} - \coth{\eta_{2}}\right),
\end{align}
with the condition $d>R_{1}+R_{2}$ ensuring that the spheres do not overlap.

The space is divided into three regions: the interior of sphere~1 ($\eta>\eta_{1}$), the interior of sphere~2 ($\eta<\eta_{2}$), and the exterior region ($\eta_{2}<\eta<\eta_{1}$), which is occupied by the surrounding dielectric medium. Each region is characterized by its own electromagnetic parameters $(\epsilon,\mu,\theta)$, with the axion field $\theta$ changing discontinuously across the spherical interfaces. This geometric formulation provides a convenient framework for solving the static electromagnetic boundary-value problem of two interacting topological insulator spheres in external fields.

\subsection{Electrodynamics of the problem}

In the static regime considered here and in the absence of free charges and currents, the electromagnetic response of the system is governed by Maxwell's equations supplemented by the axion-induced magnetoelectric coupling introduced in Sec.~\ref{Axion_ED_TIs}. Since the axion field $\theta$ is piecewise constant in the present configuration, its gradient is nonvanishing only at the spherical interfaces separating regions with different values of $\theta$. Let $\Sigma_{i}$ $(i=1,2)$ denote the interfaces between each topological insulator sphere and the surrounding dielectric medium. Across each interface, the discontinuity of the axion field gives rise to modified electromagnetic boundary conditions. In the absence of free surface charges and currents, these conditions can be written as
\begin{align}
\big[ \hat{\mathbf{n}}\cdot(\epsilon \mathbf{E}) \big]_{\Sigma}
&= \tilde{\alpha}\, \hat{\mathbf{n}}\cdot \mathbf{B}\big|_{\Sigma},
\qquad
\big[ \hat{\mathbf{n}}\times \mathbf{E} \big]_{\Sigma} = \mathbf{0},
\label{BC_E} \\[6pt]
\big[ \hat{\mathbf{n}}\cdot \mathbf{B} \big]_{\Sigma}
&= 0,
\qquad
\big[ \hat{\mathbf{n}}\times (\mathbf{B}/\mu) \big]_{\Sigma}
= - \tilde{\alpha}\, \hat{\mathbf{n}}\times \mathbf{E}\big|_{\Sigma},
\label{BC_B}
\end{align}
where $\tilde{\alpha}=\alpha(\theta/\pi)$, $\hat{\mathbf{n}}$ is the outward unit normal to the interface, and $\big[\mathbf{F}\big]_{\Sigma}=\mathbf{F}(\Sigma^{+})-\mathbf{F}(\Sigma^{-})$ denotes the discontinuity of any vector field across $\Sigma$. These relations provide a compact characterization of the full electromagnetic response at the interfaces.

To construct the solution, however, we adopt a perturbative approach in the axion coupling. Since the dimensionless parameter $\alpha=e^{2}/(4\pi\epsilon_{0}\hbar c)$ is small, the axion-induced effects can be treated as perturbative corrections to the conventional electrostatic and magnetostatic solutions. In this formulation, the effect of the axion term is incorporated through interfacial source terms proportional to $\nabla\theta$, and the modified boundary conditions need not be imposed explicitly order by order.

Accordingly, we expand the electric and magnetic fields in powers of $\alpha$ as
\begin{align}
\mathbf{E} &= \mathbf{E}^{(0)} + \mathbf{E}^{(1)} + \mathbf{E}^{(2)} + \cdots, \\
\mathbf{B} &= \mathbf{B}^{(0)} + \mathbf{B}^{(1)} + \mathbf{B}^{(2)} + \cdots,
\label{ED1}
\end{align}
where $\mathbf{E} ^{(n)}$ and $\mathbf{B} ^{(n)}$ denote contributions of order $\alpha ^{n}$. The zeroth-order fields correspond to the classical solution for two dielectric spheres in a static external electric field and satisfy the conventional Maxwell equations together with the standard electromagnetic boundary conditions at the spherical interfaces.

At first order in $\alpha$, Gauss and Ampère's laws yield
\begin{align}
\nabla \cdot ( \epsilon \mathbf{E}^{(1)} ) = -
\frac{\alpha}{\pi}\,\nabla \theta \cdot \mathbf{B}^{(0)}, \qquad \nabla \times \mathbf{B}^{(1)} =
\frac{\alpha}{\pi}\,\nabla\theta \times \mathbf{E}^{(0)},
\label{ED2}
\end{align}
while the remaining Maxwell equations retain their homogeneous form.

Introducing the first-order elecromagnetic potentials $\phi ^{(1)}$ and $\mathbf{A}^{(1)}$ as usual, and working in the Coulomb gauge $\nabla\cdot\mathbf{A}^{(1)}=0$, Eq.~\eqref{ED2} reduces to
\begin{align}
\nabla \cdot ( \epsilon \nabla \phi ^{(1)} ) =  
\frac{\alpha}{\pi}\,\nabla \theta \cdot \mathbf{B}^{(0)}, \quad \nabla ^{2} \mathbf{A}^{(1)} = -
\frac{\alpha}{\pi}\,\nabla\theta \times  \mathbf{E} ^{(0)} .
\label{ED3}
\end{align}
The solution of Eq.~\eqref{ED3} can be expressed in terms of the Green’s function as
\begin{align}
\phi ^{(1)}  (\mathbf{r}) = - 
\frac{\alpha}{\pi}
\int G _{\epsilon} (\mathbf{r},\mathbf{r}')
\,\nabla'\theta(\mathbf{r}')
\cdot \mathbf{B} ^{(0)} (\mathbf{r}')
\,\mathrm{d}^{3}\mathbf{r}' ,  \\ \mathbf{A}^{(1)}(\mathbf{r}) =   
\frac{\alpha}{\pi}
\int G _{0} (\mathbf{r},\mathbf{r}')
\,\nabla'\theta(\mathbf{r}')
\times \mathbf{E} ^{(0)} (\mathbf{r}')
\,\mathrm{d}^{3}\mathbf{r}' , 
\label{ED4}
\end{align}
where $G _{\epsilon} (\mathbf{r},\mathbf{r}')$ and $G _{0} (\mathbf{r},\mathbf{r}')$ are the Green's functions associated with the operators $- \nabla \cdot ( \epsilon \nabla \; \cdot ) $ and $- \nabla ^{2}\; \cdot $, respectively.

In the present geometry the axion field is piecewise constant: $\theta = \theta _{1}$ inside sphere 1, $\theta = \theta _{2}$ inside sphere 2, and $\theta = \theta _{m} = 0$ in the surrounding dielectric. Therefore, $\nabla \theta$ vanishes everywhere in the bulk and is supported only at the spherical interfaces $\Sigma _{1}$ and $\Sigma _{2}$, where $\theta$ is discontinuous. In distributional form one may write
\begin{align}
\nabla\theta(\mathbf{r})
=
\Delta\theta_{1}\,\delta(\Sigma_{1})\,\hat{\mathbf{n}}_{1}
+
\Delta\theta_{2}\,\delta(\Sigma_{2})\,\hat{\mathbf{n}}_{2},
\quad
\Delta\theta_{i}\equiv \theta_{m}-\theta_{i},
\label{gradtheta_general}
\end{align}
where $\hat{\mathbf{n}}_{i}$ is the outward unit normal to $\Sigma_{i}$ (pointing from the interior of sphere $i$ toward the exterior dielectric region).

In bispherical coordinates, the surfaces of the spheres are given by $\eta = \eta _{1} > 0$ and $\eta = \eta _{2} < 0$. Using $\nabla f = \hat{\boldsymbol{\eta}} \, (1/h _{\eta}) \, \partial _{\eta} f + \cdots$ with the metric factor
\begin{align}
h_{\eta}=\frac{a}{\cosh\eta-\cos\xi},
\end{align}
the interfacial delta distributions become
\begin{align}
\delta(\Sigma_{i})\,\hat{\mathbf{n}}_{i}
=
\frac{\cosh\eta-\cos\xi}{a}\,
\hat{\boldsymbol{\eta}}\,
\delta(\eta-\eta_{i}),
\qquad (i=1,2),
\label{deltaSigma_bisph}
\end{align}
so that Eq.~\eqref{gradtheta_general} yields
\begin{align}
\nabla\theta
=
\frac{\cosh\eta-\cos\xi}{a}\,
\hat{\boldsymbol{\eta}}\,
\Big[
\Delta\theta_{1}\,\delta(\eta-\eta_{1})
+\Delta\theta_{2}\,\delta(\eta-\eta_{2})
\Big].
\label{gradtheta_bisph_two_spheres}
\end{align}
Equation~\eqref{gradtheta_bisph_two_spheres} makes explicit that the axion-induced sources are localized on each spherical interface, with strengths fixed by the corresponding jumps $\Delta \theta _{i}$. Consequently, volume integrals involving $\nabla \theta$ (such as in Eq.~\eqref{ED4}) reduce to surface contributions on $\Sigma _{1}$ and $\Sigma _{2}$.

\section{Zeroth-order solution: classical two-sphere problem} \label{zeroth_order_solution}

We begin by constructing the electromagnetic fields at zeroth order in the axion coupling, $\theta = 0$. In this limit, the topological magnetoelectric response is absent and the problem reduces to the classical electrostatic response of two dielectric spheres embedded in a homogeneous dielectric medium and subjected to an externally applied uniform electric field. The zeroth-order fields provide the baseline solution upon which the axion-induced corrections are built in the subsequent sections.

Since the configuration is static and free of charges and currents, the electric field can be expressed in terms of a scalar potential $\phi ^{(0)}$ satisfying Laplace's equation in each region,
\begin{align}
\nabla^{2} \phi ^{(0)} = 0,
\end{align}
together with the standard electromagnetic boundary conditions at the spherical interfaces. The symmetry of the problem depends on the orientation of the external field relative to the line joining the centers of the spheres. Accordingly, we consider separately the two canonical configurations: (i) an external electric field parallel to the center-to-center axis, and (ii) an external electric field perpendicular to that axis.

The general solution of Laplace's equation in bispherical coordinates can be written as a superposition of separable modes adapted to the spherical geometry. A convenient basis of solutions is given by
\begin{align}
f^{\pm,\pm'}_{n m}(\eta,\xi,\phi)
=
\sqrt{\cosh\eta-\cos\xi}\;
e^{\pm\left(n+\frac{1}{2}\right)\eta}
\,P_{n}^{m}(\cos\xi)\,
e^{\pm' i m \phi},
\label{Laplace_bispherical_modes}
\end{align}
where $P_{n}^{m}$ are the associated Legendre polynomials of the first kind, with $n\in\mathbb{Z}^{+}$ and $-n\leq m\leq n$. The prefactor $\sqrt{\cosh\eta-\cos\xi}$ is the conformal factor associated with the bispherical coordinate system.

A second linearly independent set of solutions involving the associated Legendre functions of the second kind, $Q _{n} ^{m} ( \cos \xi )$, also exists. However, these functions exhibit logarithmic singularities at $\cos \xi = \pm 1$ and are therefore excluded in the present analysis. Equation~\eqref{Laplace_bispherical_modes} thus provides the most general regular solution compatible with the geometry of two nonintersecting spheres.

The expansion coefficients are fixed by imposing the boundary conditions at the spherical interfaces and by matching the far-field behavior to the externally applied uniform electric field. Further simplifications follow from the symmetry of each configuration and are discussed separately below.

\subsection{Parallel external electric field}

We first consider the case in which the externally applied electric field is parallel to the line joining the centers of the spheres, $\mathbf{E}_{0}=E_{0}\hat{\mathbf{z}}$. The corresponding external potential is $\phi_{0}=-E_{0}z$, which is odd under reflections with respect to the $xy$ plane. As a consequence, the electrostatic problem exhibits axial symmetry and all physical quantities are independent of the azimuthal angle $\phi$.

The odd parity of the external potential further implies that the induced potentials inside and outside the spheres satisfy
\begin{align}
\phi_{e}(-\eta,\xi)=-\phi_{e}(\eta,\xi), \qquad
\phi_{+}(-\eta,\xi)=-\phi_{-}(\eta,\xi),
\end{align}
so that it is sufficient to determine a single interior potential, which we choose as $\phi_{+}(\eta,\xi)$.

At the surface of each sphere, the electrostatic boundary conditions require the continuity of the potential and of the normal component of the electric displacement field,
\begin{align}
\phi_{+}(\eta_{0},\xi) &= \phi_{e}(\eta_{0},\xi), \\
\epsilon_{i}\,\partial_{\eta}\phi_{+}(\eta_{0},\xi)
&=\epsilon_{e}\,\partial_{\eta}\phi_{e}(\eta_{0},\xi),
\end{align}
where $\eta = \eta _{0}$ defines the spherical interfaces.

Guided by the general solution of Laplace’s equation in bispherical coordinates, we expand the exterior and interior potentials as
\begin{widetext}
\begin{align}
\phi_{e}(\eta,\xi)
&=(\cosh\eta-\cos\xi)^{1/2}
\sum_{n=0}^{\infty}
\left[
A_{n}\sinh(\bar{n}\eta)
-2^{3/2}E_{0}a\,\bar{n}e^{-\bar{n}\eta}
\right]
P_{n}(\cos\xi), \label{phie_parallel}\\
\phi_{+}(\eta,\xi)
&=(\cosh\eta-\cos\xi)^{1/2}
\sum_{n=0}^{\infty}
B_{n}e^{-\bar{n}\eta}
P_{n}(\cos\xi), \label{phip_parallel}
\end{align}
where $\bar{n}=n+1/2$ and the second term in Eq.~\eqref{phie_parallel} ensures the correct asymptotic behavior imposed by the external field.
\end{widetext}

Imposing continuity of the potential at $\eta=\eta_{0}$ allows the coefficients $B_{n}$ to be expressed algebraically in terms of $A_{n}$. The continuity of the normal component of the electric field then leads to a system of coupled three-term recurrence relations for the coefficients $A_{n}$, which can be written compactly as
\begin{align}
\mathcal{C}_{n-1}A_{n-1}
+\mathcal{C}_{n}A_{n}
+\mathcal{C}_{n+1}A_{n+1}
=
\mathcal{S}_{n},
\label{recurrence_parallel}
\end{align}
where the coefficients $\mathcal{C}_{n}$ and the source term $\mathcal{S}_{n}$ depend on the geometric parameter $\eta_{0}$ and on the dielectric contrast
\begin{align}
\Delta=\frac{\epsilon_{i}-\epsilon_{e}}{\epsilon_{i}+\epsilon_{e}}.
\end{align}
The explicit form of Eq.~\eqref{recurrence_parallel} is derived in Appendix~\ref{App_parallel}.

Since $|\Delta|<1$ for physical dielectrics and $e^{-\eta_{0}}<1$ for nonoverlapping spheres, the recurrence relation can be solved perturbatively in the small parameter $\Delta e^{-2n\eta_{0}}$. Writing
\begin{align}
A_{n}=A_{n}^{(0)}+A_{n}^{(1)}+A_{n}^{(2)}+\cdots,
\end{align}
one obtains a hierarchy of linear equations at successive orders.

At leading order, the solution is
\begin{align}
A_{n}^{(0)}
=
\frac{2e^{-\eta_{0}}}{3-\Delta}
\left[
n-\frac{e^{-2\eta_{0}}}{1-e^{-2\eta_{0}}}
\right],
\label{A0_parallel}
\end{align}
which corresponds to the classical electrostatic response of two dielectric spheres in a uniform external field.

The first-order correction describes the leading effect of multiple electrostatic scattering between the spheres and reads
\begin{widetext}
\begin{align}
A_{n}^{(1)}
=
\frac{2\Delta e^{-2n\eta_{0}}}{3-\Delta}
e^{-2\eta_{0}}
\left[
n
-
\frac{1+\Delta}{2}e^{-2\eta_{0}}
+
\frac{1-\Delta}{4}e^{-4\eta_{0}}
+\mathcal{O}(e^{-6\eta_{0}})
\right].
\label{A1_parallel}
\end{align}
\end{widetext}

At second order, one finds
\begin{align}
A_{n}^{(2)}
=
\Delta e^{-2n\eta_{0}}
\left[
\frac{2e^{-3\eta_{0}}}{3-\Delta}n
-
\frac{1+\Delta}{3-\Delta}e^{-5\eta_{0}}
+\mathcal{O}(e^{-7\eta_{0}})
\right],
\label{A2_parallel}
\end{align}
which accounts for higher-order multiple reflections of the induced field between the two spherical interfaces.

Together, Eqs.~\eqref{A0_parallel}-\eqref{A2_parallel} provide an accurate representation of the electrostatic potential in the regime of moderately separated spheres. These results form the basis for the calculation of the axion-induced electromagnetic response discussed in the subsequent sections.

\begin{figure}
    \centering
    \includegraphics[width=0.45\linewidth]{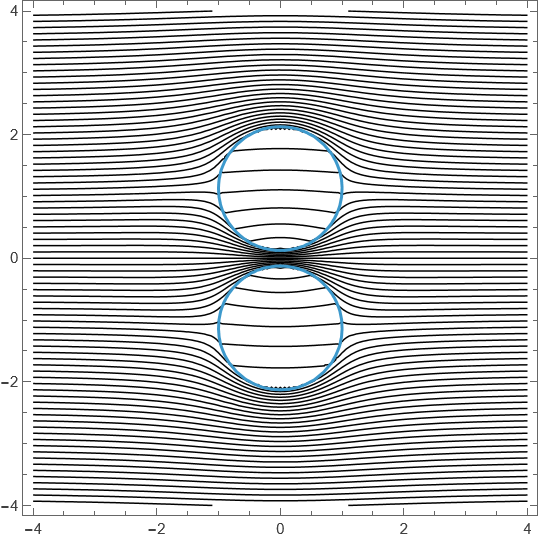} \quad \includegraphics[width=0.45\linewidth]{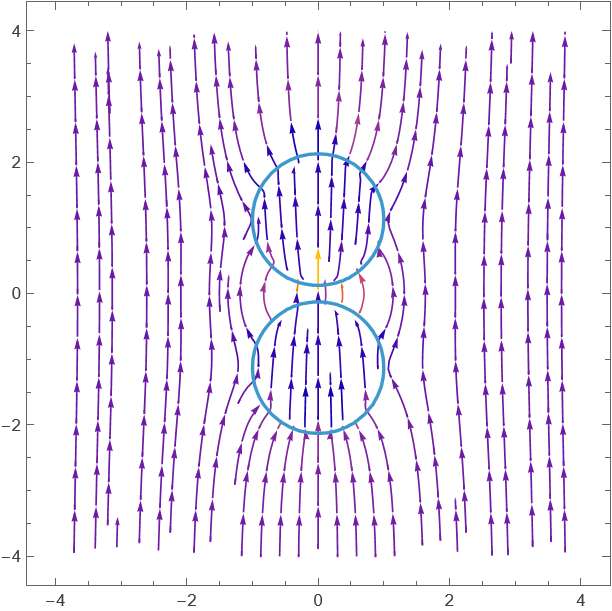} 
    \caption{Zeroth-order electrostatic configuration for two identical dielectric spheres embedded in a homogeneous dielectric medium and subjected to a uniform external electric field parallel to the center-to-center axis. The spheres are separated along the vertical $z$-axis, while the horizontal axis corresponds to the $x$-direction. Owing to the axial symmetry of the geometry, the system is invariant under azimuthal rotations around the $z$-axis. The left panel shows the equipotential lines, whereas the right panel displays the corresponding electric field lines. Far from the spheres, the field approaches the uniform external field, while near the spherical interfaces the mutual electrostatic interaction distorts both the equipotentials and the field lines, with a pronounced enhancement in the inter-sphere region.}    \label{figs_parallel_field}
\end{figure}

In Fig.~\ref{figs_parallel_field} we illustrate the zeroth-order electrostatic solution for two identical dielectric spheres separated along the center-to-center axis and subjected to a uniform external electric field oriented parallel to this axis. The left panel displays the equipotential lines, while the right panel shows the corresponding electric field lines. The distortion of both the potential contours and the field trajectories clearly reflects the mutual polarization of the spheres induced by the external field. In particular, the equipotential lines are noticeably compressed in the inter-sphere region, indicating an enhanced local electric field resulting from the electrostatic interaction between the induced multipolar moments. The electric field lines remain continuous across the dielectric interfaces but exhibit pronounced bending near the spherical surfaces, consistent with the boundary conditions imposed by the dielectric mismatch.

{Unlike the isolated single-sphere problem, for which the internal electrostatic field is exactly uniform and the equipotential surfaces inside the sphere are strictly planar, the interacting two-sphere geometry generally breaks the full spherical symmetry and induces higher-order multipolar contributions. As a consequence, the electric field inside each sphere is not exactly uniform, and the internal equipotential contours acquire a slight curvature, as can be appreciated in the present configuration. This departure from uniformity is a direct manifestation of the electrostatic coupling between the two spheres.

It is worth emphasizing, however, that this nonuniformity depends sensitively on the separation between the spheres. In the limit of large center-to-center distance, the mutual interaction becomes negligible, higher-order multipolar corrections are progressively suppressed, and each sphere asymptotically recovers the behavior of an isolated dielectric sphere embedded in a uniform external field, for which the internal electric field becomes exactly homogeneous. The configuration shown here corresponds to a moderate separation chosen to make the interaction-induced deviations from the isolated-sphere limit visually explicit.

Far from the spheres, the field lines recover the uniform external-field configuration, confirming the localized nature of the perturbation. This electrostatic solution provides the baseline configuration upon which the axion-induced magnetoelectric response is computed at leading order.

}

\subsection{Perpendicular external electric field}
\label{subsec:perpendicular}

We now consider the configuration in which the externally applied uniform electric field is perpendicular to the line joining the centers of the two spheres. Without loss of generality, we take $\mathbf{E} _{0} = E _{0} \hat{\mathbf{y}}$, so that the corresponding far-field electrostatic potential is $\phi _{0} = - E _{0} y$.

In contrast to the parallel configuration, this orientation breaks the reflection symmetry with respect to the $xy$ plane and excites angular modes with azimuthal dependence. In bispherical coordinates, the Cartesian coordinate $y$ is proportional to $\sin \xi \sin \phi / ( \cosh \eta - \cos \xi ) $, implying that the external field couples selectively to the $m=1$ sector of the bispherical harmonics. As a consequence, the electrostatic potential must be expanded in terms of the associated Legendre functions $P _{n} ^{1} ( \cos \xi )$ multiplied by $\sin \phi$.

Guided by the general solution of Laplace's equation and by the required asymptotic matching to the uniform external field, we write the potential in the exterior dielectric region as
\begin{widetext}
\begin{align}
\phi_{e}(\eta,\xi,\phi)
=
(\cosh\eta-\cos\xi)^{1/2}
\sum_{n=1}^{\infty}
\left[
C_{n}\cosh(\bar{n}\eta)
-
2^{3/2}E_{0}a\,e^{-\bar{n}\eta}
\right]
P_{n}^{1}(\cos\xi)\sin\phi,
\label{phie_perp}
\end{align}
\end{widetext}
where $\bar{n} = n + \tfrac{1}{2}$. The second term inside the brackets represents the particular solution associated with the applied field and guarantees the correct far-field behavior, while the coefficients $C _{n}$ encode the electrostatic response of the two-sphere system.

Inside the upper sphere, regularity at $\eta \to + \infty$ restricts the solution to decaying modes, and the potential can be written as
\begin{align}
\phi_{+}(\eta,\xi,\phi)
=
(\cosh\eta-\cos\xi)^{1/2}
\sum_{n=1}^{\infty}
D_{n}\,e^{-\bar{n}\eta}\,
P_{n}^{1}(\cos\xi)\sin\phi.
\label{phip_perp}
\end{align}
The potential inside the lower sphere follows by symmetry.

At the spherical interface $\eta = \eta _{0}$, continuity of the potential yields an algebraic relation between the interior and exterior coefficients,
\begin{align}
D_{n}e^{-\bar{n}\eta_{0}}
=
C_{n}\cosh(\bar{n}\eta_{0})
-
2^{3/2}E_{0}a\,e^{-\bar{n}\eta_{0}},
\label{DnCn_perp}
\end{align}
which allows the coefficients $D _{n}$ to be eliminated in favor of $C _{n}$.

The continuity of the normal component of the electric displacement field,
$\epsilon_{i}\partial_{\eta}\phi_{+}
=
\epsilon_{e}\partial_{\eta}\phi_{e}$,
then leads, after differentiation and reduction of the angular structure using standard recurrence relations of the associated Legendre functions, to a three-term difference equation for the coefficients $C _{n}$. The explicit form of this recurrence relation and its derivation are given in Appendix~\ref{App_perp}.

The resulting equation couples neighboring multipoles and depends on the geometric parameter $\eta _{0}$ and on the dielectric contrast
\begin{align}
\Delta = \frac{\epsilon _{i} - \epsilon _{e}}{\epsilon _{i} + \epsilon _{e}}.
\end{align}
For physically relevant configurations one has $|\Delta| < 1$, while the condition of nonoverlapping spheres implies $e ^{ - \eta _{0}} < 1$. Inspection of the recurrence relation shows that the coupling between different multipole orders is suppressed by factors of $\Delta e ^{- 2 n \eta _{0}}$, which naturally suggests a perturbative solution of the form
\begin{align}
C_{n}=C_{n}^{(0)}+C_{n}^{(1)}+C_{n}^{(2)}+\cdots .
\end{align}

At leading order, the recurrence relation admits a constant solution,
\begin{align}
C_{n}^{(0)}=\frac{2}{\Delta-3}\,e^{-\eta_{0}},
\label{C0_perp_main}
\end{align}
which corresponds to the dipolar response of each sphere in the presence of the transverse uniform field, neglecting multiple scattering effects.

Higher-order corrections describe successive electrostatic reflections between the two spheres. Up to second order, these corrections can be written in the compact form
\begin{align}
C_{n}^{(1)}&=
\Delta e^{-2n\eta_{0}}
\left(
\alpha^{(1)}+\frac{\beta^{(1)}}{n}
\right)
+\mathcal{O}(e^{-6\eta_{0}}),
\quad
\alpha^{(1)}=\frac{2e^{-2\eta_{0}}}{3-\Delta},
\label{C1_perp_main}\\[2mm]
C_{n}^{(2)}&=
\Delta e^{-2n\eta_{0}}
\left(
\alpha^{(2)}+\frac{\beta^{(2)}}{n}
\right),
\qquad
\alpha^{(2)}=e^{-\eta_{0}}\alpha^{(1)},
\label{C2_perp_main}
\end{align}
where the coefficients $\beta ^{(1)}$ and $\beta ^{(2)}$ are given explicitly in Appendix~\ref{App_perp}.

The perturbative solution reveals that the perpendicular field excites a well-defined hierarchy of $m = 1$ multipolar contributions, whose amplitudes are progressively suppressed by powers of $e ^{-\eta _{0}}$. This structure allows the electrostatic problem to be truncated consistently at second order for the purposes of the axion-induced analysis.

\begin{figure}
    \centering
    \includegraphics[width=0.45\linewidth]{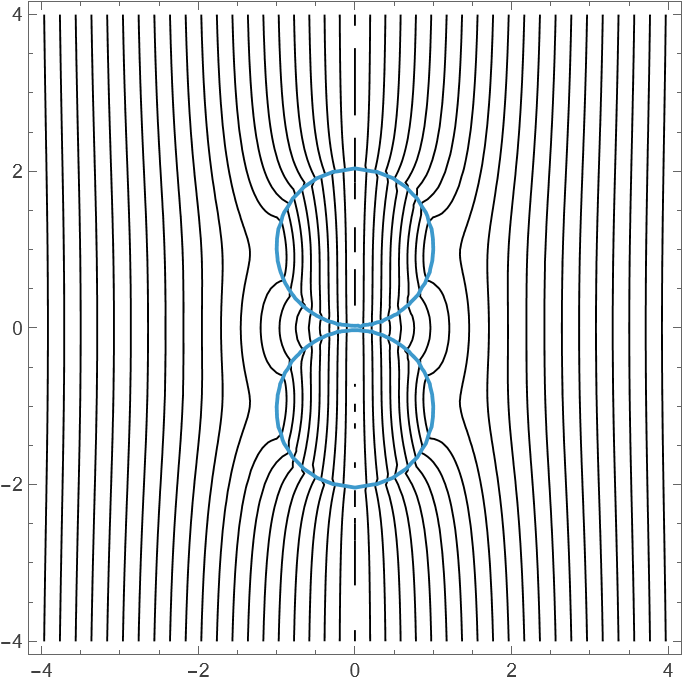} \quad \includegraphics[width=0.45\linewidth]{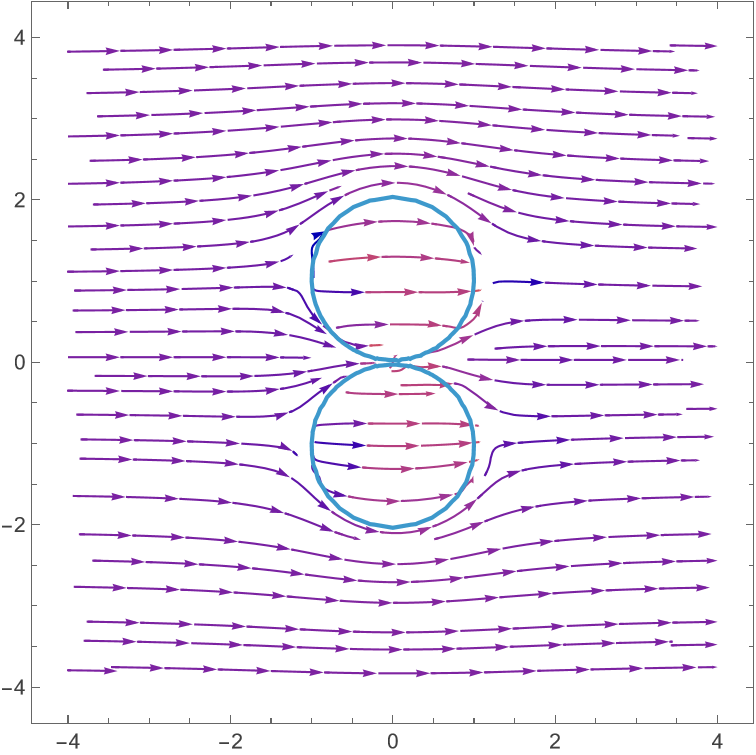} 
    \caption{Zeroth-order electrostatic configuration for two identical dielectric spheres embedded in a homogeneous dielectric medium and subjected to a uniform external electric field perpendicular to the center-to-center axis. The spheres are separated along the vertical $z$-axis, while the external field is applied along the horizontal $y$-direction. The left panel shows the equipotential lines, whereas the right panel displays the corresponding electric field lines. In contrast to the parallel configuration, the electric field does not exhibit a pronounced enhancement in the inter-sphere region; instead, the field lines are predominantly deflected sideways as they flow around the spherical interfaces. Far from the spheres, the field approaches the uniform external-field configuration, confirming the localized nature of the electrostatic perturbation.}   \label{figs_perpendicular_field}
\end{figure}

In Fig.~\ref{figs_perpendicular_field} we show the zeroth-order electrostatic solution for two identical dielectric spheres when the external electric field is oriented perpendicular to the center-to-center axis. As in the parallel configuration, the left panel displays the equipotential lines, while the right panel shows the corresponding electric field lines. In this geometry, the external field is directed along the horizontal $y$-axis, whereas the spheres are separated along the vertical $z$-axis, leading to a qualitatively different distortion pattern compared to the parallel case.

The equipotential lines exhibit a lateral squeezing around each sphere, with a pronounced asymmetry between the regions facing the external field and those aligned along the inter-sphere axis. Unlike the parallel configuration, the interstitial region between the two spheres does not display a strong enhancement of the electric field. Instead, the field lines are predominantly deflected sideways, flowing around the spheres with only weak mutual focusing. This reflects the reduced electrostatic coupling between the induced multipoles in the perpendicular geometry.

The electric field lines remain continuous across the dielectric interfaces and bend smoothly around the spherical surfaces, consistent with the electrostatic boundary conditions. {Unlike the isolated single-sphere problem, where the internal electric field is exactly uniform and the equipotential surfaces inside the sphere are strictly planar, the interacting two-sphere geometry generally induces a spatially nonuniform internal field due to the breaking of full spherical symmetry by the neighboring sphere. In the perpendicular configuration, however, this effect is comparatively weaker than in the parallel case, since the electrostatic interaction between the induced multipoles is less pronounced. As a result, the internal distortion remains relatively mild, although the equipotential contours still exhibit a measurable departure from exact planarity.

As in the parallel geometry, this interaction-induced nonuniformity gradually disappears as the center-to-center separation increases. In the large-distance limit, the electrostatic coupling between the spheres becomes negligible, the higher-order multipolar corrections are suppressed, and each sphere asymptotically recovers the exact isolated-sphere behavior, characterized by a homogeneous internal electric field in the presence of the external uniform field.

Far from the spheres, the field lines recover the uniform external-field configuration along the $y$-direction. The absence of strong field amplification in the region between the spheres highlights the crucial role of geometry in controlling the electrostatic interaction. This perpendicular configuration therefore provides a contrasting baseline to the parallel case and serves as a complementary reference for assessing the geometry-dependent axion-induced magnetoelectric response discussed in the subsequent sections.

}

\subsection{Mapping to the magnetostatic problem}

The electrostatic analysis developed in the previous sections admits a direct and exact mapping to the corresponding magnetostatic problem of two permeable spheres in an external magnetic field. This mapping follows from the formal equivalence between electrostatics in linear dielectric media and magnetostatics in linear magnetic media in the absence of free sources.

We consider two spheres of magnetic permeability $\mu_i$ embedded in a medium of permeability $\mu_e$, subjected to a uniform external magnetic field $\mathbf{H}_0$, oriented either parallel or perpendicular to the center-to-center axis. In the static, source-free regime, the magnetic fields satisfy
\begin{align}
\nabla\times\mathbf{H} &= \mathbf{0},\\
\nabla\cdot\mathbf{B} &= 0,
\end{align}
with the constitutive relation $\mathbf{B}=\mu\,\mathbf{H}$. Since $\nabla\times\mathbf{H}=0$, the magnetic field can be expressed in terms of a scalar magnetic potential,
\begin{align}
\mathbf{H} = -\nabla\psi,
\end{align}
which satisfies Laplace’s equation in each region of space,
\begin{align}
\nabla^{2}\psi = 0.
\end{align}

The boundary conditions at each spherical interface are the continuity of the magnetic scalar potential and of the normal component of the magnetic induction,
\begin{align}
\psi_i = \psi_e, \qquad
\mu_i\,\partial_n \psi_i = \mu_e\,\partial_n \psi_e,
\end{align}
which are formally identical to the electrostatic boundary conditions for the electric potential $\phi$ upon the replacement $\epsilon \rightarrow \mu$.

As a consequence, the complete magnetostatic solution can be obtained from the electrostatic one through the substitutions
\begin{align}
\phi \;\rightarrow\; \psi, \qquad
E_0 \;\rightarrow\; H_0, \qquad
\epsilon_i \;\rightarrow\; \mu_i, \qquad
\epsilon_e \;\rightarrow\; \mu_e,
\end{align}
with no further modifications to the functional form of the solution. In particular, all bispherical mode expansions, recurrence relations, and perturbative solutions derived for the electrostatic problem remain valid under this mapping.

This correspondence allows us to treat electric and magnetic responses on the same footing, and will be exploited in the following sections when discussing the axion-induced magnetoelectric coupling of the topological insulator spheres.

\begin{figure}
    \centering
    \includegraphics[width=0.45\linewidth]{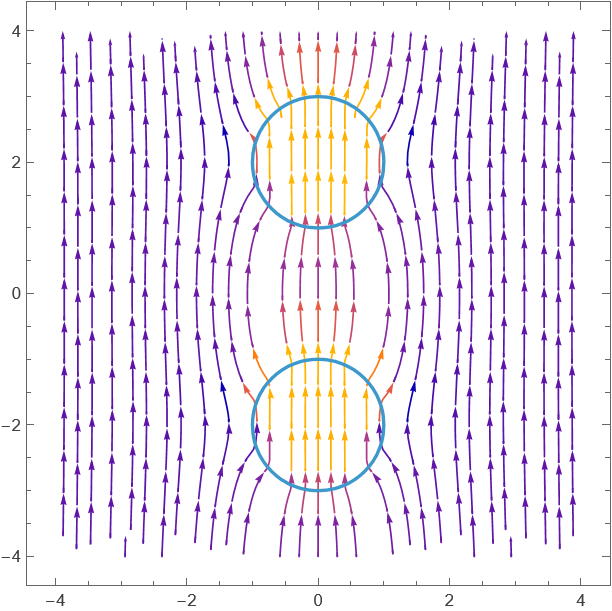} \quad \includegraphics[width=0.45\linewidth]{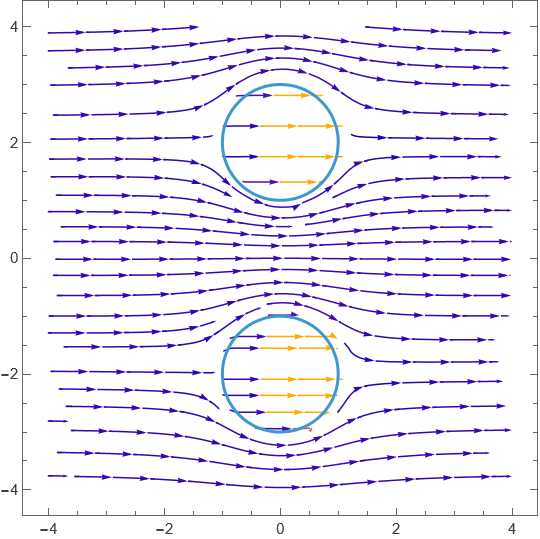} 
    \caption{Magnetostatic field lines for two identical permeable spheres embedded in a homogeneous medium and subjected to a uniform external magnetic field. The spheres are separated along the $z$-axis. The left panel corresponds to a magnetic field applied parallel to the center-to-center axis, while the right panel shows the perpendicular configuration. In the parallel case, the field lines concentrate in the inter-sphere region, whereas for the perpendicular orientation they are mainly deflected around the spheres. Far from the spheres, the magnetic field approaches the uniform external configuration.}  \label{figs_magnetic_field}
\end{figure}

To illustrate the magnetostatic problem we show in Fig.~\ref{figs_magnetic_field} the magnetic field lines for two identical permeable spheres embedded in a homogeneous medium and subjected to a uniform external magnetic field. As in the electrostatic case, the spheres are separated along the $z$-axis, while the external magnetic field is applied either parallel (left panel) or perpendicular (right panel) to this axis.

In the parallel configuration (left panel), the magnetic field lines are strongly distorted in the region between the two spheres, exhibiting a clear concentration along the center-to-center axis. This behavior reflects the constructive coupling between the induced magnetic dipoles, which align with the external field and reinforce each other in the inter-sphere region. Near the spherical interfaces, the field lines bend smoothly and intersect the surfaces in a manner consistent with the magnetostatic boundary conditions imposed by the permeability contrast. Far from the spheres, the magnetic field approaches the uniform external configuration, confirming the localized nature of the induced perturbation.

In contrast, when the external magnetic field is oriented perpendicular to the center-to-center axis (right panel), the field lines are predominantly deflected sideways as they flow around the spheres. In this geometry, the mutual interaction between the induced magnetic moments is weaker, and no pronounced enhancement of the magnetic field is observed in the region between the spheres. Instead, the distortion remains localized near each sphere, highlighting the strong dependence of the magnetostatic response on the relative orientation between the external field and the sphere separation axis.

\section{Axion-induced electromagnetic response} \label{induced_magnetic_field_section}

We now turn to the electromagnetic response induced by the axion term. Using the classical electrostatic solutions obtained above as the zeroth-order input, we treat the topological magnetoelectric coupling perturbatively in the fine-structure constant $\alpha$. In the present geometry the axion field is piecewise constant, so that the axion-induced sources are confined to the spherical interfaces and generate interfacial electromagnetic responses. We analyze the leading axion-induced contribution for different orientations of the externally applied electric field.

\subsection{Parallel configuration: axion-induced magnetic response}

We now turn to the leading axion-induced correction in the configuration $\mathbf{E}_{0}=E_{0}\hat{\mathbf{z}}$. Since the zeroth-order electrostatic potential $\phi^{(0)}(\eta,\xi)$ is axisymmetric, all zeroth-order quantities are independent of the azimuthal angle $\phi$. In addition, the axion field $\theta$ is piecewise constant and its gradient is supported exclusively on the spherical interfaces, as made explicit by Eq.~\eqref{gradtheta_bisph_two_spheres}. As a result, the axion-induced source terms entering the first-order Maxwell equations are localized at the interfaces and the Green-function representation \eqref{ED4} reduces to a purely interfacial contribution.

Using $\mathbf{E}^{(0)}=-\nabla\phi^{(0)}$ and the fact that $\partial_{\phi}\phi^{(0)}=0$, the axion-induced current density entering Amp\`ere's law is purely azimuthal. For two identical spheres whose surfaces are located at $\eta = \pm \eta _{0}$, one finds
\begin{widetext}
\begin{align}
\nabla\theta\times\nabla\phi^{(0)}(\eta,\xi)
=
-\frac{(\cosh\eta-\cos\xi)^{2}}{a^{2}}\,
\theta\,
\hat{\boldsymbol{\phi}}\,
\big[\delta(\eta-\eta_{0})+\delta(\eta+\eta_{0})\big]\,
\partial_{\xi}\phi^{(0)}(\eta,\xi),
\label{ED6_main}
\end{align}
\end{widetext}
where $\hat{\boldsymbol{\phi}}=-\hat{\mathbf{x}}\sin\phi+\hat{\mathbf{y}}\cos\phi$. Equation~\eqref{ED6_main} shows that the axion-induced surface current on each sphere circulates around the symmetry axis, which implies that the induced vector potential can be chosen to be purely azimuthal,
\begin{align}
\mathbf{A}^{(1)}(\eta,\xi,\phi)
=
A^{(1)}_{\phi}(\eta,\xi)\,\hat{\boldsymbol{\phi}},
\qquad
\partial_{\phi}A^{(1)}_{\phi}=0 .
\label{Aphi_ansatz}
\end{align}

Substituting Eq.~\eqref{ED6_main} into the Green-function representation \eqref{ED4} yields
    \begin{align}
\mathbf{A}^{(1)}(\mathbf{r})
=
-\frac{\alpha\theta}{\pi}
\int d^{3}\mathbf{r}'\;
\hat{\boldsymbol{\phi}}'\,
G _{0} (\mathbf{r},\mathbf{r}')
\frac{a\sin\xi'}{\cosh\eta'-\cos\xi'}\,
\big[\delta(\eta'-\eta_{0})+\delta(\eta'+\eta_{0})\big]\,
\partial_{\xi'}\phi^{(0)}(\eta',\xi') , 
\label{ED7_main}
\end{align}
where
\begin{align}\label{FG8}
    G _{0} (\mathbf{r},\mathbf{r}') = - \frac{(\cosh{\eta}-\cos{\xi})^{1/2}(\cosh{\eta\prime}-\cos{\xi^{\prime}})^{1/2}}{a}\sum\limits_{n,m}\frac{1}{2\bar{n}}e^{-\bar{n}\abs{\eta-\eta^{\prime}}}Y_n^m\qty(\xi,\phi)Y_n^{m\, \ast }\qty(\xi^{\prime},\phi^{\prime}) 
\end{align}
Performing the $\eta'$ integration collapses the volume integral into the sum of two surface integrals, one over each spherical interface. The remaining angular integrals can be evaluated by expanding the Green function $G _{0} (\mathbf{r},\mathbf{r}')$ in bispherical harmonics and exploiting their orthogonality properties.

Carrying out this procedure, as detailed in Appendix~\ref{App_axion_parallel}, leads to a compact series representation for the azimuthal component of the vector potential,
\begin{align}
A^{(1)}_{\phi}(\eta,\xi)
=
-\frac{\alpha\theta}{\pi}\,
(\cosh\eta-\cos\xi)^{1/2}
\sum_{n=0}^{\infty}
\mathcal{K}_{n}(\eta;\eta_{0})\,
P_{n}^{1}(\cos\xi),
\label{Aphi_series_main}
\end{align}
where the kernel $\mathcal{K}_{n}$ depends on the interface locations $\pm\eta_{0}$ and on the zeroth-order electrostatic coefficients through the combination $A_{n}\sinh(\bar{n}\eta_{0})-2^{3/2}E_{0}a\,\bar{n}e^{-\bar{n}\eta_{0}}$, with $\bar{n}=n+1/2$. The explicit form of $\mathcal{K}_{n}$ is given in Appendix~\ref{App_axion_parallel}.

Since $\mathbf{A}^{(1)}$ is purely azimuthal, the induced magnetic field $\mathbf{B}^{(1)}=\nabla\times\mathbf{A}^{(1)}$ has only $\eta$ and $\xi$ components. In bispherical coordinates it can be written as
\begin{align}
\mathbf{B}^{(1)}
=
\frac{(\cosh\eta-\cos\xi)^{2}}{a^{2}\sin\xi}
\Big[
\hat{\boldsymbol{\eta}}\,\partial_{\xi}
-
\hat{\boldsymbol{\xi}}\,\partial_{\eta}
\Big]
\left(
\frac{\sin\xi}{\cosh\eta-\cos\xi}\,
A^{(1)}_{\phi}(\eta,\xi)
\right),
\label{B_from_Aphi_main}
\end{align}
in agreement with the expression used in the derivation. The magnetic field lines in the meridional $(\eta,\xi)$ plane follow from the tangency condition
$d\boldsymbol{\ell}\times\mathbf{B}^{(1)}=\mathbf{0}$, with
$d\boldsymbol{\ell}=\frac{a}{\cosh\eta-\cos\xi}
(d\eta\,\hat{\boldsymbol{\eta}}+d\xi\,\hat{\boldsymbol{\xi}})$. This yields the first integral
\begin{align}
\frac{\sin\xi}{\cosh\eta-\cos\xi}\,
A^{(1)}_{\phi}(\eta,\xi)
=
\text{const.},
\label{fieldlines_main}
\end{align}
showing that the axion-induced magnetic streamlines are given by the level sets of the scalar function $(\sin\xi/(\cosh\eta-\cos\xi))A^{(1)}_{\phi}(\eta,\xi)$.

\begin{figure}
    \centering
    \includegraphics[width=0.45\linewidth]{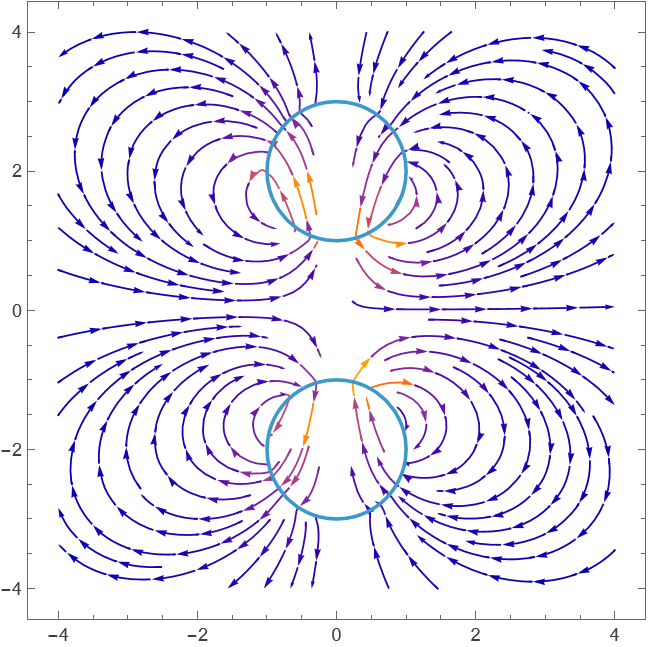} \qquad \includegraphics[width=0.16\linewidth]{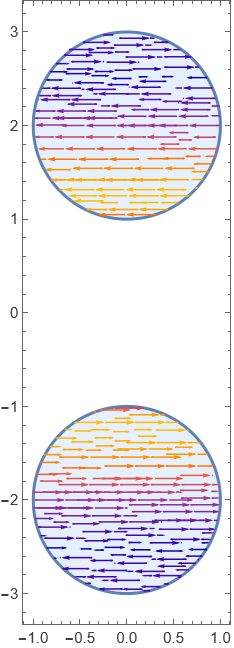} 
    \caption{Axion-induced magnetic response for two topological insulator spheres subjected to an external electric field parallel to the center-to-center axis. The left panel shows the magnetic field lines generated by the axion-induced surface currents, while the right panel displays the corresponding induced surface current density. The current is purely interfacial and corresponds to a Hall-type response localized on the spherical surfaces, giving rise to a magnetostatic field with a clear multipolar structure. }
 \label{figs_induced_magnetic_field_parallel}
\end{figure}

We now turn to the axion-induced magnetic response for the configuration in which the external electric field is applied parallel to the center-to-center axis of the two spheres. Figure~\ref{figs_induced_magnetic_field_parallel}, left panel, shows the resulting magnetic field lines generated by the topological magnetoelectric effect, while the right panel displays the corresponding induced surface current density on the spheres.

The induced current is entirely localized at the spherical interfaces and corresponds to a Hall-type surface current generated by the axion term. Its spatial distribution is strongly inhomogeneous and concentrated near the polar regions of each sphere, where the normal component of the electric field is largest. The opposite orientation of the current patterns on the two spheres reflects the symmetry of the electrostatic background and the relative orientation of the induced surface Hall responses.

The magnetic field lines form closed loops characteristic of localized current distributions and exhibit a clear multipolar structure. For each individual sphere, the dominant contribution is dipolar, with an effective magnetic moment aligned parallel to the external electric field. This behavior is consistent with the interpretation of the axion term as inducing a magnetization proportional to the applied electric field. In the two-sphere configuration, however, the superposition of the individual responses leads to a nontrivial field topology, with additional quadrupolar-like features arising from the interaction between the spheres.

In particular, the magnetic field lines are strongly distorted in the region between the spheres, where the induced magnetic dipoles couple constructively along the symmetry axis. This interaction enhances the magnetic field in the inter-sphere region and gives rise to a characteristic four-lobe pattern in the surrounding space. Far from the spheres, the magnetic field decays rapidly, confirming the localized nature of the axion-induced response and its interpretation in terms of effective multipolar sources confined to the interfaces.

These results provide a direct visualization of how the surface Hall currents generated by the axion term give rise to geometry-dependent magnetostatic fields, and they highlight the close connection between the electrostatic background, the induced interfacial currents, and the resulting magnetic multipole structure.

\subsection{Perpendicular configuration: axion-induced magnetic response}

We now consider the axion-induced electromagnetic response for an externally applied electric field perpendicular to the center-to-center axis of the two spheres. In this configuration, the zeroth-order electrostatic potential $\phi ^{(0)} ( \eta , \xi , \phi ) $ obtained in Sec.~\ref{subsec:perpendicular} exhibits an explicit dependence on the azimuthal angle $\phi$ and is dominated by $m = 1$ bispherical harmonics. As a result, the reduced axial symmetry of the problem leads to a more complex structure of the axion-induced fields than in the parallel case.

At first order in the axion coupling, the vector potential satisfies
Eq.~\eqref{ED3},
\begin{align}
\nabla^{2}\mathbf{A}^{(1)}
=
-\frac{\alpha}{\pi}\,
\nabla\theta \times \nabla\phi^{(0)},
\label{EDP_main1}
\end{align}
where the gradient of the axion field $\nabla\theta$ is localized at the spherical interfaces, as given explicitly in Eq.~\eqref{gradtheta_bisph_two_spheres}. Evaluating the cross product with the zeroth-order electric field, $\mathbf{E}^{(0)}=-\nabla\phi^{(0)}$, one finds that the axion-induced source contains both azimuthal and polar contributions, reflecting the lack of axial symmetry in the perpendicular geometry.

Substitution of Eq.~\eqref{EDP_main1} into the Green-function representation \eqref{ED4} reduces the volume integral to surface contributions on the two interfaces. Expanding the Green's function in bispherical harmonics and projecting onto the appropriate angular modes, the first-order vector potential can be written as a superposition of Cartesian components,
\begin{align}
\mathbf{A}^{(1)}(\eta,\xi,\phi)
=
A_x^{(1)}(\eta,\xi,\phi)\,\hat{\mathbf{x}}
+
A_y^{(1)}(\eta,\xi,\phi)\,\hat{\mathbf{y}}
+
A_z^{(1)}(\eta,\xi)\,\hat{\mathbf{z}},
\label{EDP_main2}
\end{align}
where $A_x^{(1)}$ and $A_y^{(1)}$ involve angular harmonics with $m=0,\pm2$, while $A_z^{(1)}$ is associated with $m=\pm1$ modes.

After performing the angular integrations, the components of the vector potential can be expressed as rapidly convergent series in terms of the zeroth-order electrostatic coefficients $C_n$,
\begin{align}
A_{i}^{(1)}(\eta,\xi,\phi)
=
-\frac{\alpha\theta}{2\pi}
\sum_{n=0}^{\infty}
\Big[
C_n\cosh(\bar{n}\eta_0)
-
2^{3/2}E_0 a\,e^{-\bar{n}\eta_0}
\Big]
\,\mathcal{F}_{i,n}(\eta,\xi,\phi),
\qquad i=x,y,z,
\label{EDP_main3}
\end{align}
where the functions $\mathcal{F}_{i,n}$ encode the angular dependence and the geometry of the two interfaces. Their explicit expressions are given in Appendix~\ref{App_axion_perp}.

The axion-induced magnetic field follows from $\mathbf{B}^{(1)}=\nabla\times\mathbf{A}^{(1)}$. In contrast to the parallel configuration, all components of $\mathbf{B}^{(1)}$ are generally nonzero in the perpendicular case, reflecting the reduced symmetry of the problem. Nevertheless, the magnetic response remains entirely interfacial in origin and is fully determined by the zeroth-order electrostatic solution.

For moderate separations between the spheres ($e^{-\eta_0}\ll1$), the series in Eq.~\eqref{EDP_main3} converges rapidly, and the leading contributions provide an accurate description of the axion-induced magnetic response in the perpendicular configuration.

\begin{figure}
    \centering
    \includegraphics[width=0.45\linewidth]{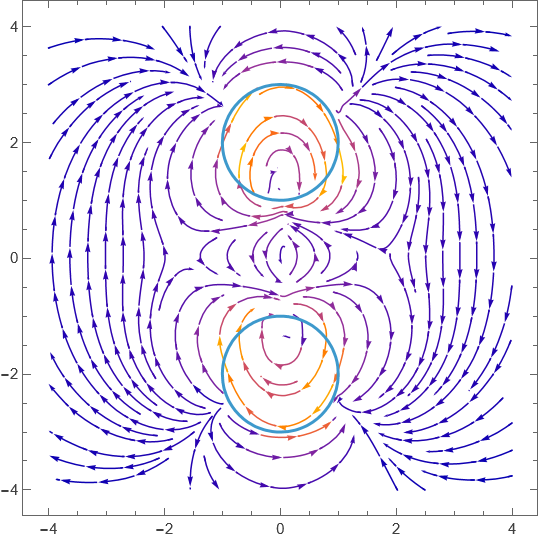} \quad
    \includegraphics[width=0.45\linewidth]{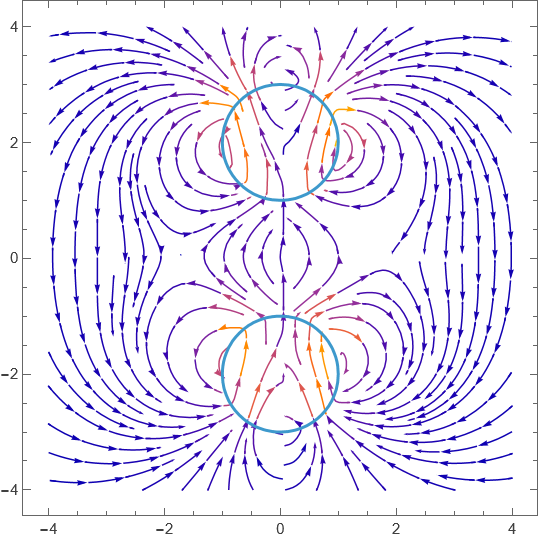} \quad
    \includegraphics[width=0.16\linewidth]{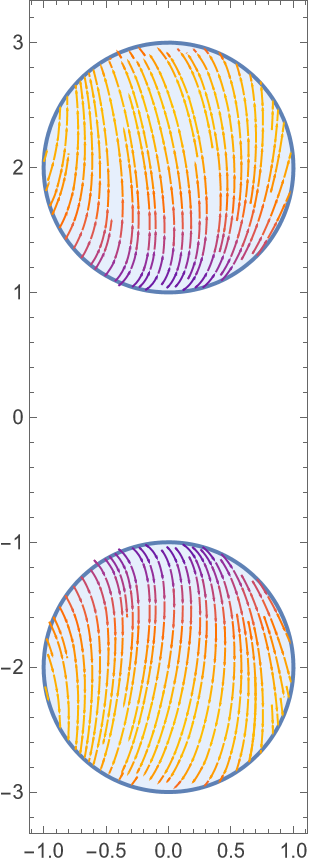} \quad \includegraphics[width=0.16\linewidth]{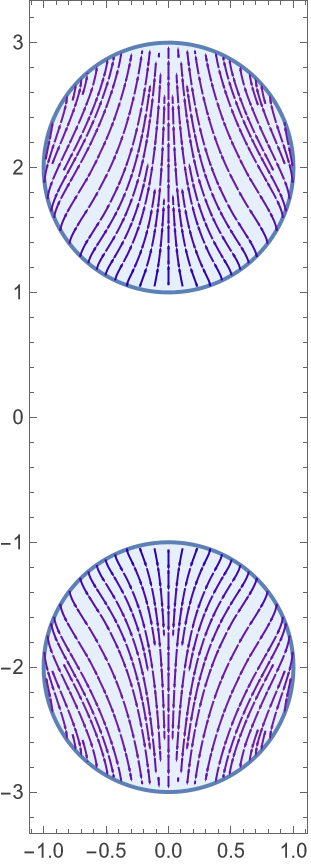} 
    \caption{Axion-induced magnetic field and surface Hall current for the perpendicular configuration.  The upper panels display the magnetic field lines generated by the axion-induced surface currents: the upper-left panel corresponds to a transverse cross section, while the upper-right panel shows the field lines in the plane $\phi = \pi/10$. The lower panels show the corresponding transverse sections of the surface Hall current on the spheres; the color scale represents the local current magnitude. The comparison between the different cuts highlights the rotation and redistribution of both the magnetic field and the surface current, reflecting the breaking of azimuthal symmetry produced by the orientation of the external electric field.} \label{figs_induced_magnetic_field_perpendicular} 
\end{figure}

As we know, the induced magnetic field is sourced by an interfacial Hall current on each topological-insulator sphere, 
$\mathbf{J}_{\!H}\propto \hat{\mathbf{n}} \times \mathbf{E}^{(0)}$, 
driven by the tangential component of the zeroth-order electric field at the surface. 
Unlike the parallel case, the external field is not aligned with the center-to-center axis, and the response no longer exhibits full azimuthal symmetry about $z$. 
As a consequence, the surface current develops an intrinsically three-dimensional pattern.

Figure~\ref{figs_induced_magnetic_field_perpendicular} displays the resulting magnetostatic configuration together with representative transverse sections of the surface current. 
The upper panels show the magnetic field lines generated by the axion-induced Hall currents: the upper-left panel corresponds to a transverse cross section, while the upper-right panel shows the field structure in the plane $\phi=\pi/10$, thereby exposing the angular dependence of the response. 
The lower panels show transverse cuts of the surface current distribution, with the left panel corresponding to a cut in the $y$ direction and the right panel to a cut in the $x$ direction; the color scale represents the local current magnitude.

The $y$-cut reveals a largely symmetric flow on each sphere: the current lines are organized in broad meridional streams on the visible cross section, with a clear up-down reversal between the upper and lower spheres, consistent with the reflection symmetry of the two-sphere arrangement with respect to the midplane between them. 
In this section the current magnitude varies smoothly along the surface, indicating that the response is governed by the regular spatial variation of $\mathbf{E}^{(0)}$.

The $x$-cut, on the other hand, displays a systematic rotation of the current pattern relative to the $y$-section, reflecting the explicit breaking of azimuthal symmetry introduced by the lateral orientation of the external field. 
Rather than producing a strongly irregular distribution, the current remains coherent over extended surface regions, but its direction and intensity are redistributed between the two hemispheres of each sphere according to the sign of the local tangential field. 
This controlled departure from axial symmetry signals the presence of angular components with $m=\pm1$, in contrast with the purely axisymmetric structure characteristic of the parallel geometry.

The magnetic field lines shown in the upper panels directly mirror this behavior. 
In the transverse section, the field forms closed-loop structures characteristic of localized current sources, while the $\phi=\pi/10$ plane makes explicit the three-dimensional deformation of the field pattern. 
Far from the spheres, the magnetic field decays rapidly, confirming its multipolar nature and the absence of net magnetic charge. 
Together, these panels make clear that the perpendicular configuration leads to a nonaxisymmetric but smooth magnetostatic response entirely determined by the geometry of the induced surface Hall currents.

These surface Hall currents constitute the sole sources of the axion-induced magnetostatic field in the bulk. 
The nonaxisymmetric but smooth character observed in the transverse cuts anticipates a magnetic response with mixed multipolar content, whose detailed field-line topology will be presented once the corresponding $\mathbf{B}$ configuration is available.

\section{Results and Discussion} \label{results_discussion_section}

In this work we have obtained an analytically controlled description of the static electromagnetic response of two spherical topological insulators embedded in a dielectric medium and subjected to a uniform external electric field. The problem was addressed by combining an exact zeroth-order electrostatic solution in bispherical coordinates with a perturbative treatment of the axion-induced magnetoelectric coupling. This strategy allows us to clearly disentangle classical dielectric effects from genuinely topological contributions and to identify the physical mechanisms responsible for the induced magnetostatic response.

At zeroth order in the axion coupling, the electrostatic problem reduces to the classical response of two dielectric spheres in an external field. By exploiting the separability of Laplace's equation in bispherical coordinates, we obtained exact mode expansions for both parallel and perpendicular orientations of the applied field. In each case, the boundary conditions at the spherical interfaces lead to three-term recurrence relations for the expansion coefficients. These relations encode multiple electrostatic scattering between the spheres and admit a natural perturbative solution in the regime of nonoverlapping spheres, where $e^{-\eta_0}\ll 1$. Explicit solutions up to second order were constructed, providing a quantitatively accurate seed for the subsequent axion-induced analysis.

The axion-induced electromagnetic response arises at first order in the fine-structure constant and is entirely interfacial in origin. Because the axion field is piecewise constant, the induced sources are localized on the spherical boundaries and are fully determined by the zeroth-order electric field evaluated at the interfaces. This feature allows the axion-induced fields to be expressed in closed form in terms of the previously obtained electrostatic coefficients, without the need to impose modified boundary conditions order by order.

In the parallel configuration, the axial symmetry of the problem implies that the axion-induced surface currents circulate azimuthally around the center-to-center axis. As a result, the induced vector potential is purely azimuthal and the corresponding magnetic field lies in the meridional $(\eta,\xi)$ plane. The magnetic streamlines are governed by a single scalar invariant,
\begin{equation}
\frac{\sin\xi}{\cosh\eta-\cos\xi}\,A^{(1)}_\phi(\eta,\xi)=\mathrm{const.},
\end{equation}
which provides a transparent geometrical interpretation of the induced magnetostatic structure. The resulting field lines resemble closed loops linking the two spheres, reflecting the circulating nature of the axion-induced surface currents.

In contrast, the perpendicular configuration exhibits a richer angular structure due to the reduced symmetry of the zeroth-order electrostatic potential. In this case, the axion-induced sources generate vector-potential components along all three Cartesian directions. The corresponding magnetic field generally possesses nonvanishing $\eta$, $\xi$, and $\phi$ components, leading to more intricate magnetostatic patterns. Nevertheless, the response remains fully controlled by the same perturbative framework: all axion-induced fields can be written as rapidly convergent series whose amplitudes are fixed by the electrostatic coefficients $C_n$ obtained at zeroth order.

An important outcome of our analysis is the rapid convergence of both the electrostatic and axion-induced series for moderate sphere separations. In practice, retaining terms up to second order in the perturbative expansion is sufficient to capture the leading interaction effects between the spheres and to construct reliable magnetostatic field profiles. This makes the present approach particularly suitable for analytical and semi-analytical studies of finite topological systems, where purely numerical treatments often obscure the underlying physical mechanisms.

From a physical perspective, our results highlight how the topological magnetoelectric effect manifests itself in interacting finite geometries. The induced magnetic response does not arise from intrinsic magnetism, but from the coupling between the externally induced electric polarization and the topological axion term at the interfaces. The dependence of the induced fields on the orientation of the external electric field provides a clear qualitative signature of the axion electrodynamics of topological insulators, which could be exploited in {experimental probes of magnetoelectric nanoparticles, including topological and more general isotropic magnetoelectric materials.}

Finally, while the present work focuses on static fields and nonoverlapping spheres, the framework developed here can be extended in several directions. Possible generalizations include spheres with different radii or axion parameters, configurations involving external magnetic fields, and dynamical responses at finite frequency. More broadly, our results provide a concrete analytical platform for exploring axion-induced electromagnetic phenomena in interacting mesoscopic systems, bridging the gap between idealized single-particle models and realistic multi-object geometries.

\acknowledgements{J.C.G., M.I.-M. and L.M.O. was supported by the SECIHTI fellowships No. 1165841, No. 4065997 and No. 834773, respectively.   A.M.-R. acknowledges financial support by UNAM-PAPIIT project No. IG100224, UNAM-PAPIME project No. PE109226, by SECIHTI project No. CBF-2025-I-1862 and by the Marcos Moshinsky Foundation.}

\appendix

\section{Derivation and perturbative solution of the recurrence relations for the parallel configuration}
\label{App_parallel}

In this appendix we present the derivation of the recurrence relations satisfied by the expansion coefficients of the electrostatic potential in the case of an external electric field parallel to the center-to-center axis, together with their perturbative solution up to second order. The purpose is to make explicit the main steps of the calculation while avoiding unnecessary algebraic repetition.

\subsection{From boundary conditions to recurrence relations}

The electrostatic potentials inside and outside the spheres are expanded as
\begin{align}
\phi_{e}(\eta,\xi)
&=
(\cosh\eta-\cos\xi)^{1/2}
\sum_{n=0}^{\infty}
\left[
A_{n}\sinh(\bar{n}\eta)
-
2^{3/2}E_{0}a\,\bar{n}e^{-\bar{n}\eta}
\right]
P_{n}(\cos\xi),
\\
\phi_{+}(\eta,\xi)
&=
(\cosh\eta-\cos\xi)^{1/2}
\sum_{n=0}^{\infty}
B_{n}e^{-\bar{n}\eta}
P_{n}(\cos\xi),
\end{align}
with $\bar{n}=n+1/2$.

Imposing continuity of the potential at the spherical surface $\eta=\eta_0$
leads to the algebraic relation
\begin{align}
B_{n}e^{-\bar{n}\eta_0}
=
A_{n}\sinh(\bar{n}\eta_0)
-
2^{3/2}E_{0}a\,\bar{n}e^{-\bar{n}\eta_0},
\label{AppBn}
\end{align}
which allows the interior coefficients $B_n$ to be eliminated in favor of
the exterior ones $A_n$.

The continuity of the normal component of the electric displacement field,
$\epsilon_i \partial_\eta \phi_{+} = \epsilon_e \partial_\eta \phi_{e}$,
requires the computation of $\partial_\eta \phi_{e}$ and
$\partial_\eta \phi_{+}$. After differentiation and expansion in Legendre
polynomials, terms proportional to $\cos\xi\,P_n(\cos\xi)$ are reduced using
the standard recurrence relation
\begin{align}
\cos\xi\,P_n(\cos\xi)
=
\frac{n}{2n+1}P_{n-1}(\cos\xi)
+
\frac{n+1}{2n+1}P_{n+1}(\cos\xi).
\end{align}
Equating the coefficients of equal Legendre polynomials then yields a
three-term recurrence relation for $A_n$ of the form
\begin{align}
\mathcal{C}_{n-1}A_{n-1}
+
\mathcal{C}_{n}A_{n}
+
\mathcal{C}_{n+1}A_{n+1}
=
\mathcal{S}_{n},
\label{AppRecFinal}
\end{align}
where the explicit expressions of the coefficients $\mathcal{C}_n$ and the
source term $\mathcal{S}_n$ depend on the geometric parameter $\eta_0$ and
on the dielectric contrast
\begin{align}
\Delta = \frac{\epsilon_i-\epsilon_e}{\epsilon_i+\epsilon_e}.
\end{align}

\subsection{Perturbative structure of the recurrence}

For nonoverlapping spheres one has $e^{-\eta_0}<1$, while for physical
dielectrics $|\Delta|<1$. Inspection of Eq.~\eqref{AppRecFinal} shows that
the coupling between neighboring coefficients $A_{n\pm1}$ is suppressed by
factors of $e^{-2n\eta_0}$, which motivates a perturbative expansion in the
small parameter $\Delta e^{-2n\eta_0}$. We therefore write
\begin{align}
A_n = A_n^{(0)} + A_n^{(1)} + A_n^{(2)} + \cdots,
\end{align}
where $A_n^{(k)}=\mathcal{O}((\Delta e^{-2n\eta_0})^{k})$.

\subsection{Zeroth-order solution}

At leading order the recurrence relation simplifies considerably and
reduces to a linear difference equation with polynomial coefficients in
$n$. Guided by its structure, we seek a solution of the form
\begin{align}
A_n^{(0)} = \alpha \, n + \beta.
\end{align}
Substitution into the zeroth-order recurrence equation and matching the
coefficients of equal powers of $n$ yields a system of algebraic equations
for $\alpha$ and $\beta$, whose solution is
\begin{align}
A_n^{(0)}
=
\frac{2 e^{-\eta_0}}{3-\Delta}
\left[
n - \frac{e^{-2\eta_0}}{1-e^{-2\eta_0}}
\right].
\label{AppA0}
\end{align}
This expression corresponds to the classical electrostatic response of two
dielectric spheres in a uniform external field.

\subsection{First-order correction}

At first order, the recurrence relation becomes inhomogeneous, with a source
term entirely determined by $A_n^{(0)}$. Since the inhomogeneity is
proportional to $e^{-2n\eta_0}$, we adopt the ansatz
\begin{align}
A_n^{(1)} = \Delta e^{-2n\eta_0}\left(\alpha^{(1)} n + \beta^{(1)}\right).
\end{align}
Substitution into the first-order recurrence relation and comparison of the
terms proportional to $n^2$ immediately yields
\begin{align}
\alpha^{(1)} = \frac{2e^{-2\eta_0}}{3-\Delta}.
\end{align}
The remaining coefficient $\beta^{(1)}$ follows from the terms linear and
independent in $n$. Expanding consistently in powers of $e^{-\eta_0}$ one
obtains
\begin{align}
\beta^{(1)}
=
-\frac{1+\Delta}{3-\Delta} e^{-4\eta_0}
+
\frac{1-\Delta^2}{2(3-\Delta)} e^{-6\eta_0}
+
\mathcal{O}(e^{-8\eta_0}),
\end{align}
leading to
\begin{align}
A_n^{(1)}
=
\frac{2\Delta e^{-2n\eta_0}}{3-\Delta}
e^{-2\eta_0}
\left[
n
-
\frac{1+\Delta}{2} e^{-2\eta_0}
+
\frac{1-\Delta}{4} e^{-4\eta_0}
+
\mathcal{O}(e^{-6\eta_0})
\right].
\label{AppA1}
\end{align}

\subsection{Second-order correction}

The second-order recurrence relation is driven by the first-order
coefficients $A_n^{(1)}$. The same structural considerations motivate the
ansatz
\begin{align}
A_n^{(2)} = \Delta e^{-2n\eta_0}\left(\alpha^{(2)} n + \beta^{(2)}\right).
\end{align}
Matching the highest-order terms in $n$ yields
\begin{align}
\alpha^{(2)} = e^{-\eta_0}\alpha^{(1)},
\end{align}
while the remaining constant $\beta^{(2)}$ is obtained by collecting the
lower-order terms. Expanding in powers of $e^{-\eta_0}$, one finds
\begin{align}
\beta^{(2)}
=
-\frac{1+\Delta}{3-\Delta} e^{-5\eta_0}
+
\frac{1+3\Delta^2-2\Delta}{2(3-\Delta)} e^{-7\eta_0}
+
\mathcal{O}(e^{-9\eta_0}).
\end{align}
Thus,
\begin{align}
A_n^{(2)}
=
\Delta e^{-2n\eta_0}
\left[
e^{-\eta_0}\alpha^{(1)} n
-
\frac{1+\Delta}{3-\Delta} e^{-5\eta_0}
+
\mathcal{O}(e^{-7\eta_0})
\right].
\label{AppA2}
\end{align}

The perturbative expansion converges rapidly for moderate separations between the spheres, where $e^{-\eta_0}\ll1$. Higher-order corrections describe multiple electrostatic scattering between the spheres and can be obtained recursively following the same strategy. In the present work, terms up to second order provide sufficient accuracy for constructing the axion-induced electromagnetic response discussed in the main text.

\section{Derivation and perturbative solution of the recurrence relations for the perpendicular configuration}
\label{App_perp}

In this appendix we derive the recurrence relations satisfied by the expansion coefficients of the electrostatic potential in the perpendicular configuration, $\mathbf{E}_{0}=E_{0}\hat{\mathbf{y}}$, and present their perturbative solution up to second order. The goal is to make explicit the logical structure of the calculation while keeping the algebraic details to a manageable level.

\subsection{Mode selection and exterior potential}

In the perpendicular configuration the far-field electrostatic potential is
\begin{align}
\phi_{0} = -E_{0}y,
\end{align}
which, in bispherical coordinates, exhibits an explicit $\sin\phi$ angular dependence. As a consequence, only bispherical harmonics with azimuthal index $m=1$ contribute, and the angular structure is entirely captured by the combination $P_{n}^{1}(\cos\xi)\sin\phi$.

A convenient expansion for the exterior potential is therefore
\begin{align}
\phi_{e}(\eta,\xi,\phi)=
(\cosh\eta-\cos\xi)^{1/2}
\sum_{n=1}^{\infty}C_{n}\cosh(\bar{n}\eta)\,
P_{n}^{1}(\cos\xi)\sin\phi
-\frac{E_{0}a\sin\xi\sin\phi}{\cosh\eta-\cos\xi},
\label{App_l10_phi}
\end{align}
with $\bar{n}=n+1/2$. The second term represents the externally applied field.

Using the identity
\begin{align}
P_{n}^{1}(x)=(1-x^{2})^{1/2}\,\dv{P_{n}(x)}{x},
\qquad
P_{n}^{1}(\cos\xi)=\sin\xi\,P_{n}'(\cos\xi),
\end{align}
together with the standard bispherical expansion of $(\cosh\eta-\cos\xi)^{-3/2}$, the external-field contribution can be recast in the same $P_{n}^{1}$ basis. This leads to the compact representation
\begin{align}
\phi_{e}(\eta,\xi,\phi)
=
(\cosh\eta-\cos\xi)^{1/2}
\sum_{n=1}^{\infty}
\left[
C_{n}\cosh(\bar{n}\eta)
-
2^{3/2}E_{0}a\,e^{-\bar{n}\eta}
\right]
P_{n}^{1}(\cos\xi)\sin\phi.
\label{App_phie_perp}
\end{align}

\subsection{Interior potential and continuity of the potential}

Inside the upper sphere ($\eta>\eta_{0}$), regularity as $\eta\to+\infty$ requires exponentially decaying modes. The interior potential is therefore expanded as
\begin{align}
\phi_{+}(\eta,\xi,\phi)
=
(\cosh\eta-\cos\xi)^{1/2}
\sum_{n=1}^{\infty}D_{n}e^{-\bar{n}\eta}
P_{n}^{1}(\cos\xi)\sin\phi.
\label{App_phip_perp}
\end{align}

Continuity of the electrostatic potential at the spherical interface $\eta=\eta_{0}$ yields the algebraic relation
\begin{align}
D_{n}e^{-\bar{n}\eta_{0}}
=
C_{n}\cosh(\bar{n}\eta_{0})
-
2^{3/2}E_{0}a\,e^{-\bar{n}\eta_{0}},
\label{App_l22_phi}
\end{align}
which allows the interior coefficients $D_{n}$ to be eliminated in favor of the exterior ones $C_{n}$.

\subsection{Continuity of the normal displacement and recurrence relation}

The second boundary condition enforces continuity of the normal component of the electric displacement field,
\begin{align}
\epsilon_{i}\,\partial_{\eta}\phi_{+}(\eta_{0},\xi,\phi)
=
\epsilon_{e}\,\partial_{\eta}\phi_{e}(\eta_{0},\xi,\phi).
\end{align}
Upon differentiation, the resulting expressions contain terms proportional to $P_{n}^{1}(\cos\xi)\sin\phi$ as well as to $\cos\xi\,P_{n}^{1}(\cos\xi)\sin\phi$. The latter are reduced to the $P_{n\pm1}^{1}$ basis using the standard recurrence relations for associated Legendre functions.

After eliminating $D_{n}$ using Eq.~\eqref{App_l22_phi} and equating coefficients of identical angular harmonics, one obtains a three-term recurrence relation for the coefficients $C_{n}$. Introducing the dielectric contrast
\begin{align}
\Delta=\frac{\epsilon_{i}-\epsilon_{e}}{\epsilon_{i}+\epsilon_{e}},
\end{align}
the final result can be written as
\begin{align}
(n-1)\Big(e^{\eta_{0}}+\Delta e^{-2(n-1)\eta_{0}}\Big)C_{n-1}
+\Big[\Delta\sinh\eta_{0}-(2n+1)\cosh\eta_{0}
-n\Delta e^{-2n\eta_{0}}-(n+1)\Delta e^{-2(n+1)\eta_{0}}\Big]C_{n}
\nonumber\\
+(n+2)\Big(e^{-\eta_{0}}+\Delta e^{-2(n+2)\eta_{0}}\Big)C_{n+1}
=
e^{-2\eta_{0}}-1.
\label{App_rec_perp_full}
\end{align}

\subsection{Perturbative structure of the recurrence}

For nonoverlapping spheres one has $e^{-\eta_{0}}<1$, while for physical dielectrics $|\Delta|<1$. Inspection of Eq.~\eqref{App_rec_perp_full} shows that terms coupling different multipoles are suppressed by factors of $\Delta e^{-2n\eta_{0}}$. This motivates a perturbative expansion of the form
\begin{align}
C_{n}=C_{n}^{(0)}+C_{n}^{(1)}+C_{n}^{(2)}+\cdots,
\end{align}
where $C_{n}^{(k)}=\mathcal{O}((\Delta e^{-2n\eta_{0}})^{k})$.

\subsection{Zeroth-order solution}

At leading order all $\Delta e^{-2n\eta_{0}}$ terms are neglected, and the recurrence reduces to
\begin{align}
(n-1)e^{\eta_{0}}C_{n-1}^{(0)}
+\Big[\Delta\sinh\eta_{0}-(2n+1)\cosh\eta_{0}\Big]C_{n}^{(0)}
+(n+2)e^{-\eta_{0}}C_{n+1}^{(0)}
=
e^{-2\eta_{0}}-1.
\label{App_rec_perp_0}
\end{align}
Since the right-hand side is independent of $n$, a constant solution $C_{n}^{(0)}=C^{(0)}$ is sufficient, yielding
\begin{align}
C_{n}^{(0)}=\frac{2}{\Delta-3}\,e^{-\eta_{0}}.
\end{align}

\subsection{First-order correction}

At first order the recurrence becomes inhomogeneous, with a source determined entirely by $C_{n}^{(0)}$. Guided by the structure of the suppressed terms, we adopt the ansatz
\begin{align}
C_{n}^{(1)}=\Delta e^{-2n\eta_{0}}
\left(\alpha^{(1)}+\frac{\beta^{(1)}}{n}\right)
+\mathcal{O}(e^{-6\eta_{0}}).
\end{align}
Matching the leading contributions in $n$ yields
\begin{align}
\alpha^{(1)}=\frac{2e^{-2\eta_{0}}}{3-\Delta},
\end{align}
while the subleading terms give
\begin{align}
\beta^{(1)}=-\frac{\Delta-1}{3-\Delta}e^{-4\eta_{0}}
+\mathcal{O}(e^{-6\eta_{0}}).
\end{align}

\subsection{Second-order correction}

The second-order recurrence is driven by the first-order solution. Using the same structural ansatz,
\begin{align}
C_{n}^{(2)}=\Delta e^{-2n\eta_{0}}
\left(\alpha^{(2)}+\frac{\beta^{(2)}}{n}\right),
\end{align}
one finds from the leading-order balance
\begin{align}
\alpha^{(2)}=e^{-\eta_{0}}\alpha^{(1)},
\end{align}
and from the subleading terms
\begin{align}
\beta^{(2)}=\frac{1-\Delta}{3-\Delta}e^{-5\eta_{0}}
+\mathcal{O}(e^{-7\eta_{0}}).
\end{align}

The perturbative expansion converges rapidly for moderate separations ($e^{-\eta_{0}}\ll1$), and terms up to second order are sufficient for the construction of the axion-induced electromagnetic response discussed in the main text.

\section{Axion-induced vector potential in the parallel configuration}
\label{App_axion_parallel}

In this appendix we provide the intermediate steps leading from the Green-function representation \eqref{ED4} to an explicit series expression for the first-order vector potential in the parallel configuration $\mathbf{E}_{0}=E_{0}\hat{\mathbf{z}}$. Throughout we assume two identical spheres with interfaces at $\eta=\pm\eta_{0}$ and a uniform axion parameter $\theta$ inside each sphere (with $\theta=0$ in the exterior medium). The generalization to two different interfaces $\eta=\eta_{1}>0$ and $\eta=\eta_{2}<0$ follows by the replacements indicated at the end of the appendix.

\subsection{Reduction of the source to the interfaces}

Since the zeroth-order solution is axisymmetric, $\phi^{(0)}=\phi^{(0)}(\eta,\xi)$ and $\partial_{\phi}\phi^{(0)}=0$. Using the bispherical expression for $\nabla\theta$ in Eq.~\eqref{gradtheta_bisph_two_spheres} (with $\eta_{1}=\eta_{0}$ and $\eta_{2}=-\eta_{0}$) one finds that the source term in Amp\`ere's law is purely azimuthal,
\begin{align}
\nabla\theta\times\nabla\phi^{(0)}(\eta,\xi)
=
-\frac{(\cosh\eta-\cos\xi)^{2}}{a^{2}}\,
\theta\,
\hat{\boldsymbol{\phi}}\,
\big[\delta(\eta-\eta_{0})+\delta(\eta+\eta_{0})\big]\,
\partial_{\xi}\phi^{(0)}(\eta,\xi),
\label{App_ED6}
\end{align}
which is equivalent to Eq.~\eqref{ED6_main}. Therefore the first-order vector potential may be chosen as
\begin{align}
\mathbf{A}^{(1)}(\eta,\xi,\phi)=A^{(1)}_{\phi}(\eta,\xi)\,\hat{\boldsymbol{\phi}},
\qquad
\partial_{\phi}A^{(1)}_{\phi}=0,
\label{App_Aphi_ansatz}
\end{align}
and Eq.~\eqref{ED4} reduces to the surface-supported integral
\begin{widetext}
    \begin{align}
\mathbf{A}^{(1)}(\mathbf{r})
=
-\frac{\alpha\theta}{\pi}
\int d^{3}\mathbf{r}'\;
G_{0}(\mathbf{r},\mathbf{r}')\,
\hat{\boldsymbol{\phi}}'\,
\frac{a\sin\xi'}{\cosh\eta'-\cos\xi'}\,
\big[\delta(\eta'-\eta_{0})+\delta(\eta'+\eta_{0})\big]\,
\partial_{\xi'}\phi^{(0)}(\eta',\xi') .
\label{App_ED7}
\end{align}
\end{widetext}
Performing the $\eta'$ integration collapses the volume integral into the sum of two surface integrals over the angular variables $(\xi',\phi')$ evaluated at $\eta'=\eta_{0}$ and $\eta'=-\eta_{0}$.

\subsection{Derivative of the zeroth-order potential at the interfaces}

For the parallel configuration the exterior potential is expanded as in the main text,
\begin{widetext}
    \begin{align}
\phi^{(0)}_{e}(\eta,\xi)
&=(\cosh\eta-\cos\xi)^{1/2}
\sum_{n=0}^{\infty}
\left[
A_{n}\sinh(\bar{n}\eta)
-2^{3/2}E_{0}a\,\bar{n}e^{-\bar{n}\eta}
\right]
P_{n}(\cos\xi),
\qquad
\bar{n}=n+\frac{1}{2}.
\label{App_phi0_parallel}
\end{align}
\end{widetext}
Since $\partial_{\xi}P_{n}(\cos\xi)=-\sin\xi\,P_{n}'(\cos\xi)$ and $P_{n}^{1}(\cos\xi)=\sin\xi\,P_{n}'(\cos\xi)$, we have the useful identity
\begin{align}
\partial_{\xi}P_{n}(\cos\xi)=-P_{n}^{1}(\cos\xi).
\label{App_dxiPn}
\end{align}
Differentiating Eq.~\eqref{App_phi0_parallel} with respect to $\xi$ gives
\begin{widetext}
    \begin{align}
\partial_{\xi}\phi^{(0)}_{e}(\eta,\xi)
&=
\frac{\sin\xi}{2(\cosh\eta-\cos\xi)^{1/2}}
\sum_{n=0}^{\infty}
\left[
A_{n}\sinh(\bar{n}\eta)
-2^{3/2}E_{0}a\,\bar{n}e^{-\bar{n}\eta}
\right]
P_{n}(\cos\xi)
\nonumber\\
&\hspace{1.7cm}
-(\cosh\eta-\cos\xi)^{1/2}
\sum_{n=0}^{\infty}
\left[
A_{n}\sinh(\bar{n}\eta)
-2^{3/2}E_{0}a\,\bar{n}e^{-\bar{n}\eta}
\right]
P_{n}^{1}(\cos\xi).
\label{App_dxi_phi0}
\end{align}
\end{widetext}
Evaluating Eq.~\eqref{App_dxi_phi0} at the interfaces $\eta=\pm\eta_{0}$ provides the explicit angular dependence entering the surface integrals in Eq.~\eqref{App_ED7}.

\subsection{Bispherical expansion of the Green function and angular projection} \label{bispherical_subsect}

We use the standard bispherical-harmonic expansion of the free-space Green function $G_{0}(\mathbf{r},\mathbf{r}')$,
    \begin{align}
G_{0}(\mathbf{r},\mathbf{r}')
=
(\cosh\eta-\cos\xi)^{1/2}(\cosh\eta'-\cos\xi')^{1/2}
\sum_{l=0}^{\infty}\sum_{m=-l}^{l}
g_{l}(\eta,\eta')\,Y_{l}^{m}(\xi,\phi)\,Y_{l}^{m}(\xi',\phi')^{*},
\label{App_G0_bisph}
\end{align}
where $g _{l}(\eta,\eta ') = - \frac{1}{2\bar{n} a}e^{-\bar{n}\abs{\eta-\eta^{\prime}}}$ is the radial kernel (symmetric in its arguments) and $Y_{l}^{m}$ are the usual spherical harmonics on the $(\xi,\phi)$ sphere. Substituting Eq.~\eqref{App_G0_bisph} into Eq.~\eqref{App_ED7} yields

\begin{align}
\mathbf{A}^{(1)}(\mathbf{r})
&=
-\frac{\alpha\theta}{\pi}
(\cosh\eta-\cos\xi)^{1/2}
\sum_{l=0}^{\infty}\sum_{m=-l}^{l}
Y_{l}^{m}(\xi,\phi)
\int d\Omega'\;
\hat{\boldsymbol{\phi}}'\,
Y_{l}^{m}(\xi',\phi')^{*}\,
\sin\xi'\,
\mathcal{I}(\eta';\xi')
\Big[
g_{l}(\eta,\eta_{0})-g_{l}(\eta,-\eta_{0})
\Big],
\label{App_A1_after_eta}
\end{align}
where $d\Omega'=\sin\xi'\,d\xi'\,d\phi'$ and we defined the interface combination
\begin{align}
\mathcal{I}(\eta';\xi')
\equiv
\left.
\frac{a}{\cosh\eta'-\cos\xi'}\,
\partial_{\xi'}\phi^{(0)}(\eta',\xi')
\right|_{\eta'=\pm\eta_{0}} .
\label{App_I_def}
\end{align}
The difference $g_{l}(\eta,\eta_{0})-g_{l}(\eta,-\eta_{0})$ originates from the relative sign between the two $\delta$-supported contributions in Eq.~\eqref{App_ED7} and makes the odd parity with respect to $\eta\to-\eta$ manifest.

Because $\hat{\boldsymbol{\phi}}'$ carries an azimuthal dependence $\propto e^{\pm i\phi'}$, the $\phi'$ integral in Eq.~\eqref{App_A1_after_eta} projects only the $m=\pm1$ harmonics. Using the standard relations
\begin{align}
\sin\phi'
=
\frac{e^{i\phi'}-e^{-i\phi'}}{2i},
\qquad
\cos\phi'
=
\frac{e^{i\phi'}+e^{-i\phi'}}{2},
\label{App_trig_exp}
\end{align}
and the explicit $\phi'$ dependence of $Y_{l}^{m}(\xi',\phi')\propto e ^{im\phi'}$, one finds that all contributions with $m\neq\pm1$ vanish by orthogonality. As a result, $\mathbf{A}^{(1)}$ is purely azimuthal and can be written as in Eq.~\eqref{App_Aphi_ansatz}.

\subsection{Final series for $A ^{(1)} _{\phi}$}

Collecting the surviving $m=\pm1$ contributions and using the relation between $Y_{l}^{\pm1}$ and $P_{l}^{1}(\cos\xi)$, the angular projection yields a compact bispherical series of the form
\begin{align}
A ^{(1)}_{\phi}(\eta,\xi)
=
-\frac{\alpha\theta}{\pi}
(\cosh\eta-\cos\xi)^{1/2}
\sum_{n=0}^{\infty}
\Big[
A_{n}\sinh(\bar{n}\eta_{0})
-2^{3/2}E_{0}a\,\bar{n}e^{-\bar{n}\eta_{0}}
\Big]\,
\Big[g_{n}(\eta,\eta_{0})-g_{n}(\eta,-\eta_{0})\Big]\,
P_{n}^{1}(\cos\xi),
\label{App_Aphi_final_simple}
\end{align}
where the kernel index has been relabeled $l\to n$ for notational
consistency with the electrostatic expansion and we used the identity \eqref{App_dxiPn} to express the interfacial derivative in the $P_{n}^{1}$ basis. Equation~\eqref{App_Aphi_final_simple} is the expression quoted in the main text, with the kernel $g_{n}$ inherited from the Green-function expansion \eqref{App_G0_bisph}. Further refinements (e.g. rewriting the $\sin\xi\,P_{n}$ term in Eq.~\eqref{App_dxi_phi0} as $n\pm1$ combinations) are possible using standard Legendre identities; we do not pursue them here since Eq.~\eqref{App_Aphi_final_simple} is already suitable for numerical evaluation once $A_{n}$ is known.

\subsection{Induced magnetic field and streamline invariant}

Given $A^{(1)}_{\phi}(\eta,\xi)$, the first-order magnetic field follows from $\mathbf{B}^{(1)}=\nabla\times\mathbf{A}^{(1)}$. For an axisymmetric purely azimuthal potential, the curl in bispherical coordinates reduces to
\begin{align}
\mathbf{B}^{(1)}
=
\frac{(\cosh\eta-\cos\xi)^{2}}{a^{2}\sin\xi}
\Big[
\hat{\boldsymbol{\eta}}\,\partial_{\xi}
-
\hat{\boldsymbol{\xi}}\,\partial_{\eta}
\Big]
\left(
\frac{\sin\xi}{\cosh\eta-\cos\xi}\,A^{(1)}_{\phi}(\eta,\xi)
\right),
\label{App_B_from_Aphi}
\end{align}
in agreement with the expression used in the main text. The field lines in the meridional $(\eta,\xi)$ plane satisfy $d\boldsymbol{\ell}\times \mathbf{B}^{(1)}=\mathbf{0}$ with $d\boldsymbol{\ell}=\frac{a}{\cosh\eta-\cos\xi}(d\eta\,\hat{\boldsymbol{\eta}} +d\xi\,\hat{\boldsymbol{\xi}})$, yielding the streamline invariant
\begin{align}
\frac{\sin\xi}{\cosh\eta-\cos\xi}\,A^{(1)}_{\phi}(\eta,\xi)=\mathrm{const.}
\label{App_streamlines}
\end{align}
Thus, the level sets of the scalar function $\frac{\sin\xi}{\cosh\eta-\cos\xi}A^{(1)}_{\phi}$ provide the axion-induced magnetic field lines around the spheres.

\subsection{Generalization to two distinct spheres}

If the spheres are not identical, their interfaces are located at $\eta=\eta_{1}>0$ and $\eta=\eta_{2}<0$, and the axion jumps are $\Delta\theta_{i}=\theta_{m}-\theta_{i}$. In this case, Eqs.~\eqref{App_ED6} and \eqref{App_ED7} generalize by the replacement
\begin{align}
\theta\big[\delta(\eta-\eta_{0})+\delta(\eta+\eta_{0})\big]
\;\longrightarrow\;
\Delta\theta_{1}\,\delta(\eta-\eta_{1})
+
\Delta\theta_{2}\,\delta(\eta-\eta_{2}),
\end{align}
and Eq.~\eqref{App_Aphi_final_simple} becomes the sum of two interface contributions,
\begin{align}
A^{(1)}_{\phi}(\eta,\xi)
=
-\frac{\alpha}{\pi}
(\cosh\eta-\cos\xi)^{1/2}
\sum_{n=0}^{\infty}
\Big\{
\Delta\theta_{1}\,\mathcal{Q}_{n}(\eta_{1})\,g_{n}(\eta,\eta_{1})
+
\Delta\theta_{2}\,\mathcal{Q}_{n}(\eta_{2})\,g_{n}(\eta,\eta_{2})
\Big\}\,
P_{n}^{1}(\cos\xi),
\end{align}
where $\mathcal{Q}_{n}(\eta_{i})$ denotes the interfacial combination of the zeroth-order coefficients evaluated at $\eta=\eta_{i}$ (the obvious analogue of the bracket in Eq.~\eqref{App_Aphi_final_simple}).

\section{Axion-induced vector potential in the perpendicular configuration} \label{App_axion_perp}

Here we present the detailed derivation of the axion-induced vector potential in the perpendicular configuration, $\mathbf{E}_{0}=E_{0}\hat{\mathbf{y}}$. The purpose is to make explicit the intermediate steps leading from the Green-function representation \eqref{ED4} to the series expressions for the Cartesian components of $\mathbf{A}^{(1)}$ quoted in the main text.

Throughout this appendix we use the distributional expression for $\nabla\theta$ given in Eq.~\eqref{gradtheta_bisph_two_spheres}, and the zeroth-order electrostatic potential $\phi^{(0)}$ obtained in Sec.~\ref{induced_magnetic_field_section}.

\subsection{Structure of the axion-induced source}

At first order in the axion coupling, the vector potential satisfies
\begin{align}
\nabla^{2}\mathbf{A}^{(1)}
=
-\frac{\alpha}{\pi}\,
\nabla\theta \times \nabla\phi^{(0)} .
\label{App_EDP1}
\end{align}
Using the bispherical-coordinate expression for the gradient operator and the fact that $\nabla\theta$ is purely normal to the interfaces, the source term can be written as
\begin{align}
\nabla\theta \times \nabla\phi^{(0)}
=
-\frac{(\cosh\eta-\cos\xi)^{2}}{a^{2}}
\Big[
\delta(\eta-\eta_{0})+\delta(\eta+\eta_{0})
\Big]
\Big[
\hat{\boldsymbol{\phi}}\partial_{\xi}
-
\hat{\boldsymbol{\xi}}\frac{1}{\sinh\eta}\partial_{\phi}
\Big]\phi^{(0)} ,
\label{App_EDP2}
\end{align}
where $\phi^{(0)}(\eta,\xi,\phi)$ is the perpendicular zeroth-order potential expanded in $P_{n}^{1}(\cos\xi)\sin\phi$ modes.

Evaluating the derivatives at $\eta=\pm\eta_{0}$ yields
\begin{align}
\Big[
\hat{\boldsymbol{\phi}}\partial_{\xi}
-
\hat{\boldsymbol{\xi}}\frac{1}{\sinh\eta}\partial_{\phi}
\Big]\phi^{(0)}(\pm\eta_{0},\xi,\phi)
&=
(\cosh\eta_{0}-\cos\xi)^{-1/2}
\sum_{n=1}^{\infty}
\Big[
C_{n}\cosh(\bar{n}\eta_{0})
-2^{3/2}E_{0}a\,e^{-\bar{n}\eta_{0}}
\Big]
\nonumber\\
&\quad\times
\Big[
\hat{\boldsymbol{\phi}}
\Big(
\partial_{\xi}
+\frac{\sin\xi}{2(\cosh\eta_{0}-\cos\xi)}
\Big)
-
\hat{\boldsymbol{\xi}}\frac{1}{\sin\xi}\partial_{\phi}
\Big]
P_{n}^{1}(\cos\xi)\sin\phi .
\label{App_EDP3}
\end{align}

\subsection{Cartesian decomposition of the source}

To facilitate the angular integrations, it is convenient to rewrite the
operator acting on the angular functions in terms of Cartesian components.
Using the explicit expressions of the bispherical basis vectors and
identifying the angular-momentum operators, one finds
\begin{align}
\hat{\boldsymbol{\phi}}\partial_{\xi}
-
(\cos\xi\cos\phi\,\hat{\mathbf{x}}
+\cos\xi\sin\phi\,\hat{\mathbf{y}})
\frac{1}{\sin\xi}\partial_{\phi}
=
\hat{\mathbf{x}}\,iL_{x}
+
\hat{\mathbf{y}}\,iL_{y},
\label{App_EDP4}
\end{align}
so that the full operator becomes
\begin{align}
\hat{\boldsymbol{\phi}}
\Big(
\partial_{\xi}
+\frac{\sin\xi}{2(\cosh\eta_{0}-\cos\xi)}
\Big)
-
\hat{\boldsymbol{\xi}}\frac{1}{\sin\xi}\partial_{\phi}
=
\hat{\mathbf{x}}\mathcal{O}_{x}
+
\hat{\mathbf{y}}\mathcal{O}_{y}
+
\hat{\mathbf{z}}\mathcal{O}_{z},
\label{App_EDP5}
\end{align}
with explicit differential operators $\mathcal{O}_{i}$ acting on the angular
functions.

\subsection{Green-function expansion and angular projections}

Substituting Eq.~\eqref{App_EDP3} into the Green-function representation
\eqref{ED4}, the $\eta'$ integration collapses onto the two spherical
interfaces. Expanding the Green’s function in spherical harmonics and using
\begin{align}
iL_{x}=\frac{i}{2}(L_{+}+L_{-}),
\qquad
iL_{y}=\frac{1}{2}(L_{+}-L_{-}),
\qquad
L_{\pm}Y_{l}^{m}
=
\sqrt{(l\mp m)(l\pm m+1)}\,Y_{l}^{m\pm1},
\end{align}
together with the identities
\begin{align}
P_{n}^{1}(\cos\xi)\sin\phi
=
-\sqrt{\frac{4\pi(n+1)!}{2\bar{n}(n-1)!}}
\frac{i}{2}
\big(
Y_{n}^{1}+Y_{n}^{-1}
\big),
\end{align}
the angular integrations can be performed explicitly by orthogonality of the
spherical harmonics.

As a result, the three Cartesian components of the vector potential are
obtained as convergent series,
\begin{align}
A_{x}^{(1)}(\eta,\xi,\phi)
&=
-\frac{\alpha\theta}{2\pi}
\sum_{n=0}^{\infty}
\Big[
C_{n}\cosh(\bar{n}\eta_{0})
-2^{3/2}E_{0}a\,e^{-\bar{n}\eta_{0}}
\Big]
\Big[
g_{n}(\eta,\eta_{0})+g_{n}(\eta,-\eta_{0})
\Big]
\Big[
P_{n}^{2}(\cos\xi)\cos2\phi
+n(n+1)P_{n}(\cos\xi)
\Big]
\nonumber\\
&\quad
+\;\text{neighbor-mode contributions},
\label{App_Ax_final}
\\[1ex]
A_{y}^{(1)}(\eta,\xi,\phi)
&=
-\frac{\alpha\theta}{2\pi}
\sum_{n=0}^{\infty}
\Big[
C_{n}\cosh(\bar{n}\eta_{0})
-2^{3/2}E_{0}a\,e^{-\bar{n}\eta_{0}}
\Big]
\Big[
g_{n}(\eta,\eta_{0})+g_{n}(\eta,-\eta_{0})
\Big]
P_{n}^{2}(\cos\xi)\sin2\phi
\nonumber\\
&\quad
+\;\text{neighbor-mode contributions},
\label{App_Ay_final}
\\[1ex]
A_{z}^{(1)}(\eta,\xi)
&=
-\frac{\alpha\theta}{2\pi}
\sum_{n=0}^{\infty}
4\bar{n}
\Big[
\frac{\epsilon_{e}\sinh(\bar{n}\eta_{0})
+\epsilon_{i}\cosh(\bar{n}\eta_{0})}{\epsilon_{e}-\epsilon_{i}}\,C_{n}
-
2^{3/2}E_{0}a\,e^{-\bar{n}\eta_{0}}
\Big]
\Big[
g_{n}(\eta,\eta_{0})-g_{n}(\eta,-\eta_{0})
\Big]
P_{n}^{1}(\cos\xi).
\label{App_Az_final}
\end{align}

Here $g_{n}(\eta,\eta_{0})$ denotes the radial Green-function kernel defined
in Sec.~\ref{bispherical_subsect}.

\bibliography{refs.bib}

@article{thouless_1982, 
    author = {Thouless, D. J. and Kohmoto, M. and Nightingale, M. P. and den Nijs, M.}, 
    title = {Quantized {H}all Conductance in a Two-Dimensional Periodic Potential}, 
    journal = {Phys. Rev. Lett.}, 
    volume = {49}, 
    pages = {405-408}, 
    year = {1982}, 
    doi = {10.1103/PhysRevLett.49.405} 
}

@article{hasan_kane_2010, 
    author = {Hasan, M. Z. and Kane, C. L.}, 
    title = {Colloquium: Topological insulators}, 
    journal = {Rev. Mod. Phys.}, 
    volume = {82}, 
    pages = {3045-3067}, 
    year = {2010}, 
    doi = {10.1103/RevModPhys.82.3045} 
}

@article{PhysRevD.94.085019,
  title = {Electromagnetic description of three-dimensional time-reversal invariant ponderable topological insulators},
  author = {Mart\'{\i}n-Ruiz, A. and Cambiaso, M. and Urrutia, L. F.},
  journal = {Phys. Rev. D},
  volume = {94},
  issue = {8},
  pages = {085019},
  numpages = {19},
  year = {2016},
  month = {Oct},
  publisher = {American Physical Society},
  doi = {10.1103/PhysRevD.94.085019},
  url = {https://link.aps.org/doi/10.1103/PhysRevD.94.085019}
}

@article{qi_zhang_2011, 
    author = {Qi, X.-L. and Zhang, S.-C.}, 
    title = {Topological insulators and superconductors}, 
    journal = {Rev. Mod. Phys.}, 
    volume = {83}, 
    pages = {1057-1110}, 
    year = {2011}, 
    doi = {10.1103/RevModPhys.83.1057} 
}

@article{fu_kane_mele_2007, 
    author = {Fu, L. and Kane, C. L. and Mele, E. J.}, 
    title = {Topological Insulators in Three Dimensions}, 
    journal = {Phys. Rev. Lett.}, 
    volume = {98}, 
    pages = {106803}, 
    year = {2007}, 
    doi = {10.1103/PhysRevLett.98.106803} 
}

@article{moore_balents_2007, 
    author = {Moore, J. E. and Balents, L.}, 
    title = {Topological invariants of time-reversal-invariant band structures}, 
    journal = {Phys. Rev. B}, 
    volume = {75}, 
    pages = {121306}, 
    year = {2007}, 
    doi = {10.1103/PhysRevB.75.121306} 
}

@article{roy_2009, 
    author = {Roy, R.}, 
    title = {Topological phases and the quantum spin {H}all effect in three dimensions}, 
    journal = {Phys. Rev. B}, 
    volume = {79}, 
    pages = {195322}, 
    year = {2009}, 
    doi = {10.1103/PhysRevB.79.195322} 
}

@article{okamoto_imura_2014,
  author  = {Okamoto, Mayuko and Takane, Yositake and Imura, Ken-Ichiro},
  title   = {One-dimensional topological insulator: a model for studying finite-size effects in topological insulator thin films},
  journal = {Phys. Rev. B},
  volume  = {89},
  pages   = {125425},
  year    = {2014},
  doi     = {10.1103/PhysRevB.89.125425}
}

@article{governale_bhandari_2020,
  author  = {Governale, Michele and Bhandari, Bibek B. and Taddei, Fabio and Imura, Ken-Ichiro and Z{\"u}licke, Ulrich},
  title   = {Finite-size effects in cylindrical topological insulators},
  journal = {New J. Phys.},
  volume  = {22},
  pages   = {063055},
  year    = {2020},
  doi     = {10.1088/1367-2630/ab90d3}
}

@article{cook_nielsen_2023,
  author  = {Cook, Ashley M. and Nielsen, Anne E. B.},
  title   = {Finite-size topology},
  journal = {Phys. Rev. B},
  volume  = {108},
  pages   = {045144},
  year    = {2023},
  doi     = {10.1103/PhysRevB.108.045144}
}

@article{qi_hughes_zhang_2008, 
    author = {Qi, X.-L. and Hughes, T. L. and Zhang, S.-C.}, 
    title = {Topological field theory of time-reversal invariant insulators}, 
    journal = {Phys. Rev. B}, 
    volume = {78}, 
    pages = {195424}, 
    year = {2008}, 
    doi = {10.1103/PhysRevB.78.195424} 
}

@article{essin_magnetoelectric_2009, 
    author = {Essin, A. M. and Moore, J. E. and Vanderbilt, D.}, 
    title = {Magnetoelectric polarizability and axion electrodynamics in crystalline insulators}, 
    journal = {Phys. Rev. Lett.}, 
    volume = {102}, 
    pages = {146805}, 
    year = {2009}, 
    doi = {10.1103/PhysRevLett.102.146805} 
}

@article{qi_monopole_2009, 
    author = {Qi, X.-L. and Li, R. and Zang, J. and Zhang, S.-C.},
    title = {Inducing a magnetic monopole with topological surface states}, 
    journal = {Science}, 
    volume = {323}, 
    pages = {1184-1187}, 
    year = {2009}, 
    doi = {10.1126/science.1167747} 
}

@article{PhysRevLett.103.171601,
  title = {Electric-Magnetic Duality and Topological Insulators},
  author = {Karch, A.},
  journal = {Phys. Rev. Lett.},
  volume = {103},
  issue = {17},
  pages = {171601},
  numpages = {4},
  year = {2009},
  month = {Oct},
  publisher = {American Physical Society},
  doi = {10.1103/PhysRevLett.103.171601},
  url = {https://link.aps.org/doi/10.1103/PhysRevLett.103.171601}
}

@article{PhysRevD.92.125015,
  title = {Green's function approach to {C}hern-{S}imons extended electrodynamics: An effective theory describing topological insulators},
  author = {Mart\'{\i}n-Ruiz, A. and Cambiaso, M. and Urrutia, L. F.},
  journal = {Phys. Rev. D},
  volume = {92},
  issue = {12},
  pages = {125015},
  numpages = {12},
  year = {2015},
  month = {Dec},
  publisher = {American Physical Society},
  doi = {10.1103/PhysRevD.92.125015},
  url = {https://link.aps.org/doi/10.1103/PhysRevD.92.125015}
}

@article{PhysRevLett.106.020403,
  title = {Tunable {C}asimir Repulsion with Three-Dimensional Topological Insulators},
  author = {Grushin, Adolfo G. and Cortijo, Alberto},
  journal = {Phys. Rev. Lett.},
  volume = {106},
  issue = {2},
  pages = {020403},
  numpages = {4},
  year = {2011},
  month = {Jan},
  publisher = {American Physical Society},
  doi = {10.1103/PhysRevLett.106.020403},
  url = {https://link.aps.org/doi/10.1103/PhysRevLett.106.020403}
}

@article{PhysRevB.84.045119,
  title = {Effect of finite temperature and uniaxial anisotropy on the {C}asimir effect with three-dimensional topological insulators},
  author = {Grushin, Adolfo G. and Rodriguez-Lopez, Pablo and Cortijo, Alberto},
  journal = {Phys. Rev. B},
  volume = {84},
  issue = {4},
  pages = {045119},
  numpages = {15},
  year = {2011},
  month = {Jul},
  publisher = {American Physical Society},
  doi = {10.1103/PhysRevB.84.045119},
  url = {https://link.aps.org/doi/10.1103/PhysRevB.84.045119}
}

@article{Martín-Ruiz_2016,
    doi = {10.1209/0295-5075/113/60005},
    url = {https://doi.org/10.1209/0295-5075/113/60005},
    year = {2016},
    month = {apr},
    publisher = {EDP Sciences, IOP Publishing and Società Italiana di Fisica},
    volume = {113},
    number = {6},
    pages = {60005},
    author = {Martín-Ruiz, A. and Cambiaso, M. and Urrutia, L. F.},
    title = {A {G}reen's function approach to the {C}asimir effect on topological insulators with planar symmetry},
    journal = {Europhysics Letters}
}

@book{jackson_classical, 
    author = {Jackson, J. D.}, 
    title = {Classical Electrodynamics}, 
    edition = {3}, 
    publisher = {Wiley}, 
    year = {1998}, 
    address = {New York} 
}

@article{PhysRevB.13.4320,
  title = {Two dielectric Spheres in an electric field},
  author = {Goyette, A. and Navon, A.},
  journal = {Phys. Rev. B},
  volume = {13},
  issue = {10},
  pages = {4320--4327},
  numpages = {0},
  year = {1976},
  month = {May},
  publisher = {American Physical Society},
  doi = {10.1103/PhysRevB.13.4320},
  url = {https://link.aps.org/doi/10.1103/PhysRevB.13.4320}
}

@article{10.1063/1.343737,
    author = {Stoy, Richard D.},
    title = {Solution procedure for the {L}aplace equation in bispherical coordinates for two spheres in a uniform external field: Perpendicular orientation},
    journal = {Journal of Applied Physics},
    volume = {66},
    number = {10},
    pages = {5093-5095},
    year = {1989},
    month = {11},
    issn = {0021-8979},
    doi = {10.1063/1.343737},
    url = {https://doi.org/10.1063/1.343737}
}

@article{10.1098/rspa.2012.0133,
    author = {Lekner, John},
    title = {Electrostatics of two charged conducting spheres},
    journal = {Proceedings of the Royal Society A: Mathematical, Physical and Engineering Sciences},
    volume = {468},
    number = {2145},
    pages = {2829-2848},
    year = {2012},
    month = {05},
    issn = {1364-5021},
    doi = {10.1098/rspa.2012.0133},
    url = {https://doi.org/10.1098/rspa.2012.0133},
}

@article{leichner_two_spheres,
    title = {Scattering of electromagnetic waves by two equal spherical particles},
    journal = {Journal of Colloid and Interface Science},
    volume = {27},
    number = {3},
    pages = {442-457},
    year = {1968},
    issn = {0021-9797},
    doi = {https://doi.org/10.1016/0021-9797(68)90182-3},
    url = {https://www.sciencedirect.com/science/article/pii/0021979768901823},
    author = {S Levine and G.O Olaofe},
}

@article{batool_two_spheres_2022,
  title = {Scattering of electromagnetic waves by two equal spherical particles},
  author = {Batool, S. and Qureshi, I. A.},
  journal = {Optik},
  volume = {249},
  pages = {168271},
  year = {2022},
  doi = {10.1016/j.ijleo.2021.168271}
}

@article{prodan_plasmonics,
  author = {Prodan, E. and Nordlander, P.},
  title = {Plasmon hybridization in spherical nanoparticles},
  journal = {J. Chem. Phys.},
  volume = {120},
  pages = {5444--5454},
  year = {2004},
  doi = {10.1063/1.1647512}
}

@article{markel_dimer_review,
author = {V.A. Markel},
title = {Coupled-dipole Approach to Scattering of Light from a One-dimensional Periodic Dipole Structure},
journal = {Journal of Modern Optics},
volume = {40},
number = {11},
pages = {2281--2291},
year = {1993},
publisher = {Taylor \& Francis},
doi = {10.1080/09500349314552291},
URL = {https://doi.org/10.1080/09500349314552291}
}

@article{brongersma_plasmon_review,
  title = {Plasmon-induced hot carrier science and technology},
  author = {Brongersma, M. L. and Halas, N. J. and Nordlander, P.},
  journal = {Nature Nanotechnology},
  volume = {10},
  pages = {25--34},
  year = {2015},
  doi = {10.1038/nnano.2014.311}
}

@article{10.1063/1.4961091,
    author = {Derbenev, Ivan N. and Filippov, Anatoly V. and Stace, Anthony J. and Besley, Elena},
    title = {Electrostatic interactions between charged dielectric particles in an electrolyte solution},
    journal = {The Journal of Chemical Physics},
    volume = {145},
    number = {8},
    pages = {084103},
    year = {2016},
    month = {08},
    issn = {0021-9606},
    doi = {10.1063/1.4961091},
    url = {https://doi.org/10.1063/1.4961091}
}

@article{hudlet_epjb_1998,
  author  = {Hudlet, S. and Saint Jean, M. and Guthmann, C. and Berger, J.},
  title   = {Evaluation of the capacitive force between an atomic force microscopy tip and a metallic surface},
  journal = {The European Physical Journal B},
  volume  = {2},
  number  = {1},
  pages   = {5-10},
  year    = {1998},
  doi     = {10.1007/s100510050219}
}

@article{10.1063/1.4862897,
    author = {Khachatourian, Armik and Chan, Ho-Kei and Stace, Anthony J. and Bichoutskaia, Elena},
    title = {Electrostatic force between a charged sphere and a planar surface: A general solution for dielectric materials},
    journal = {The Journal of Chemical Physics},
    volume = {140},
    number = {7},
    pages = {074107},
    year = {2014},
    month = {02},
    issn = {0021-9606},
    doi = {10.1063/1.4862897},
    url = {https://doi.org/10.1063/1.4862897}
}

@ARTICLE{7862209,
  author={Banerjee, Shubho and Levy, Mason and Davis, McKenna and Wilkerson, Blake},
  journal={IEEE Transactions on Industry Applications}, 
  title={Exact and Approximate Capacitance and Force Expressions for the Electrostatic Interaction Between Two Equal-Sized Charged Conducting Spheres}, 
  year={2017},
  volume={53},
  number={3},
  pages={2455-2460},
  doi={10.1109/TIA.2017.2672744}}

@article{PhysRevA.100.042124,
  title = {Magnetoelectric effect of a conducting sphere near a planar topological insulator},
  author = {Mart\'{\i}n-Ruiz, A. and Rodr\'{\i}guez-Tzompantzi, Omar and Maze, J. R. and Urrutia, L. F.},
  journal = {Phys. Rev. A},
  volume = {100},
  issue = {4},
  pages = {042124},
  numpages = {15},
  year = {2019},
  month = {Oct},
  publisher = {American Physical Society},
  doi = {10.1103/PhysRevA.100.042124},
  url = {https://link.aps.org/doi/10.1103/PhysRevA.100.042124}
}

@article{shalaev_metamaterials_review,
	author = {Shalaev, Vladimir M. },
	da = {2007/01/01},
	doi = {10.1038/nphoton.2006.49},
	id = {Shalaev2007},
	isbn = {1749-4893},
	journal = {Nature Photonics},
	number = {1},
	pages = {41--48},
	title = {Optical negative-index metamaterials},
	ty = {JOUR},
	url = {https://doi.org/10.1038/nphoton.2006.49},
	volume = {1},
	year = {2007}}

@article{pendry_metamaterials,
  author = {Pendry, J. B.},
  title = {Negative refraction makes a perfect lens},
  journal = {Phys. Rev. Lett.},
  volume = {85},
  pages = {3966--3969},
  year = {2000},
  doi = {10.1103/PhysRevLett.85.3966}
}

@article{nordlander_hybridization,
	author = {Prodan, E. and Nordlander, P.},
	doi = {10.1063/1.1647518},
	issn = {0021-9606},
	journal = {The Journal of Chemical Physics},
	month = {03},
	number = {11},
	pages = {5444-5454},
	title = {Plasmon hybridization in spherical nanoparticles},
	url = {https://doi.org/10.1063/1.1647518},
	volume = {120},
	year = {2004}}

@article{markel_effective_media,
  title = {Introduction to the {M}axwell {G}arnett approximation: Tutorial},
  author = {Markel, V. A.},
  journal = {Journal of the Optical Society of America A},
  volume = {33},
  pages = {1244--1256},
  year = {2016},
  doi = {10.1364/JOSAA.33.001244}
}

@book{morse_feshbach, author = {Morse, P. M. and Feshbach, H.}, title = {Methods of Theoretical Physics}, publisher = {McGraw-Hill}, year = {1953}, address = {New York} }

@book{MoonSpencer1961, title = {Field Theory Handbook}, author = {Moon, Parry and Spencer, Domina Eberle}, publisher = {Springer}, address = {Berlin}, year = {1961}, isbn = {978-3-642-45858-7} }

@book{MorseFeshbach1953, title = {Methods of Theoretical Physics}, author = {Morse, Philip M. and Feshbach, Herman}, publisher = {McGraw-Hill}, address = {New York}, year = {1953}, volume = {2} }

@article{Spaldin_2008,
doi = {10.1088/0953-8984/20/43/434203},
url = {https://doi.org/10.1088/0953-8984/20/43/434203},
year = {2008},
month = {oct},
publisher = {},
volume = {20},
number = {43},
pages = {434203},
author = {Spaldin, Nicola A and Fiebig, Manfred and Mostovoy, Maxim},
title = {The toroidal moment in condensed-matter physics and its relation to the magnetoelectric
effect},
journal = {Journal of Physics: Condensed Matter}
}

@article{Malashevich_2010,
doi = {10.1088/1367-2630/12/5/053032},
url = {https://doi.org/10.1088/1367-2630/12/5/053032},
year = {2010},
month = {may},
publisher = {},
volume = {12},
number = {5},
pages = {053032},
author = {Malashevich, Andrei and Souza, Ivo and Coh, Sinisa and Vanderbilt, David},
title = {Theory of orbital magnetoelectric response},
journal = {New Journal of Physics}
}

@article{PhysRevLett.109.197203,
  title = {Linear Magnetoelectric Effect by Orbital Magnetism},
  author = {Scaramucci, A. and Bousquet, E. and Fechner, M. and Mostovoy, M. and Spaldin, N. A.},
  journal = {Phys. Rev. Lett.},
  volume = {109},
  issue = {19},
  pages = {197203},
  numpages = {5},
  year = {2012},
  month = {Nov},
  publisher = {American Physical Society},
  doi = {10.1103/PhysRevLett.109.197203},
  url = {https://link.aps.org/doi/10.1103/PhysRevLett.109.197203}
}

\end{document}